\documentclass[superscriptaddress,nofootinbib,floatfix,twocolumn]{revtex4-2}

\usepackage{tikz}
\usepackage{amssymb,amsmath}
\usepackage{qcircuit}
\usepackage{appendix}
\usepackage{hyperref}
\usepackage{subcaption}
\usepackage[normalem]{ulem}

\definecolor{lime}{HTML}{A6CE39}
\DeclareRobustCommand{\orcidicon}{%
	\begin{tikzpicture}
	\draw[lime, fill=lime] (0,0) 
	circle [radius=0.16] 
	node[white] {{\fontfamily{qag}\selectfont \tiny ID}};
	\draw[white, fill=white] (-0.0625,0.095) 
	circle [radius=0.007];
	\end{tikzpicture}
	\hspace{-2mm}
}

\foreach \x in {A, ..., Z}{%
	\expandafter\xdef\csname orcid\x\endcsname{\noexpand\href{https://orcid.org/\csname orcidauthor\x\endcsname}{\noexpand\orcidicon}}
}

\newcommand{\appropto}{\mathrel{\vcenter{
  \offinterlineskip\halign{\hfil$##$\cr
    \propto\cr\noalign{\kern2pt}\sim\cr\noalign{\kern-2pt}}}}}

\definecolor{bbared}{rgb}{0.7,0.2,0.2}
\definecolor{bbablue}{rgb}{0.2,0.2,0.5}
\definecolor{bbagreen}{rgb}{0.0, 0.5, 0.0}

\newcommand{\ket}[1]{|#1\rangle}
\newcommand{\braket}[2]{\langle #1|#2\rangle}

\begin{document}

\title{Quantum state preparation of gravitational waves}

\author{Fergus Hayes\orcidA{}}
\email{fergus.hayes@glasgow.ac.uk}
\affiliation{SUPA, School of Physics and Astronomy, University of Glasgow, Glasgow G12 8QQ, UK}

\author{Sarah Croke\orcidB{}}
\affiliation{SUPA, School of Physics and Astronomy, University of Glasgow, Glasgow G12 8QQ, UK}

\author{Chris Messenger\orcidC{}}
\affiliation{SUPA, School of Physics and Astronomy, University of Glasgow, Glasgow G12 8QQ, UK}

\author{Fiona C.~Speirits\orcidD{}}
\affiliation{SUPA, School of Physics and Astronomy, University of Glasgow, Glasgow G12 8QQ, UK}

\date{\today}

% Small tweaks to typesetting, typos etc. {eqnarray} changed to {equation} or {align} as appropriate. Some larger (sentence-length) changes left in source code comments; a couple of comments/questions left in \textcolor{red} - FCS 11/10/22

\begin{abstract}
    We detail a quantum circuit capable of efficiently encoding analytical approximations to gravitational wave signal waveforms of compact binary coalescences into the amplitudes of quantum bits using both quantum arithmetic operations and hybrid classical-quantum generative modelling.
    The gate cost of the proposed method is considered and compared to a state preparation routine for arbitrary amplitudes, where we demonstrate up to a four orders of magnitude reduction in gate cost when considering the encoding of gravitational waveforms representative of binary neutron star inspirals detectable to the Einstein telescope.
    We demonstrate through a quantum simulation, that is limited to $28$ qubits, the encoding of a second post-Newtonian inspiral waveform with a fidelity compared to the desired state of $0.995$ when using the Grover-Rudolph algorithm, or $0.979$ when using a trained quantum generative adversarial network with a significant reduction of required gates.
\end{abstract}

\maketitle

\section{Introduction}
The Laser Interferometer Gravitational-Wave Observatory (LIGO) made the first detection of a gravitational wave in September of 2015 from a binary black-hole merger dubbed GW150914~\cite{abbott2016observation}, which began the era of gravitational wave astronomy.
Since then, the LIGO-Virgo-KAGRA collaboration has catalogued over 90 additional detections~\cite{abbott2019gwtc,abbott2021gwtca,abbott2021gwtcb} including neutron star-black hole mergers~\cite{abbott2021observation} as well as the first binary neutron star merger event GW170817~\cite{abbott2017gw170817} that was detected in conjunction with electromagnetic observations~\cite{abbott2017gravitational,Abbott_2017} --- the first time that these two observation channels had coincided for a single event.
However, there is still much that gravitational wave astronomy can offer us: from the discoveries of gravitational waves from sources such as continuous gravitational waves~\cite{abbott2021all}, supernovae~\cite{abbott2020optically}, the stochastic background~\cite{abbott2021search} and cosmic strings~\cite{abbott2021constraints}; to testing our understanding of cosmology~\cite{abbott2021gravitational} and general relativity~\cite{abbott2021tests}. 
To realize these aims will require future generations of gravitational wave~\cite{maggiore2020science,2019BAAS...51g..35R,berti2006gravitational} and electromagnetic detectors~\cite{amati2018theseus} to be constructed, the observations from which will need to be processed by ever more sophisticated data analysis methods.

Quantum computing as a field has grown independently but alongside the development of gravitational wave detectors~\cite{nielsen2002quantum}.
It is well known that quantum computers offer the potential of performing some computational tasks faster than their classical counterparts, such as prime factorisation using Shor's algorithm~\cite{shor1999polynomial} and unstructured searches using Grover's algorithm~\cite{grover1997quantum}. Since the conception of the idea in the 1980's, quantum computing has now progressed to the widespread availability of noisy quantum processors. 
Repeated, real time error-correction for a single logical qubit has recently been demonstrated experimentally, providing a first step towards fault tolerant quantum computing devices~\cite{ryan2021realization,krinner2022realizing}. %While there are still many steps yet to be taken before fault tolerant quantum computing can be performed, there has been much work \sarah{On me to update here, I am dithering a bit about what to say and will come back to this when we revisit the introduction} into how data analysis can be carried out on quantum computers given both fault tolerant and noisy intermediate-scale quantum (NISQ) devices with methods such as variational quantum algorithm~\cite{cerezo2021variational}. %\textcolor{red}{Sarah: Note that the first experimental demonstrations of error-correction for a single logical qubit have already been shown and may be worth citing: https://www.nature.com/articles/s41586-022-04566-8; https://journals.aps.org/prx/abstract/10.1103/PhysRevX.11.041058. Also I would suggest to mention variational and/or hybrid algorithms rather than quantum machine learning, which is more controversial.}

Given the development of this quantum technology and the potential of quantum algorithms to speed up some computational tasks, it is fitting that astronomers can now look to see whether quantum computing can help overcome current and future data analysis challenges.
In previous work, it has been shown that the standard signal detection analysis for gravitational waves, matched filtering~\cite{allen2012findchirp}, can acquire a quadratic speed-up using Grover's algorithm~\cite{gao2022quantum}, a procedure later built upon in \cite{miyamoto2022gravitational}.
Bayesian inference of gravitational wave parameters has also been shown to acquire a polynomial speed-up~\cite{escrig2022parameter} with a quantum version of the famous Metropolis-Hastings algorithm~\cite{hastings1970monte}.

There are a number of reasons why gravitational wave data analysis is a particularly interesting case for exploring applications for quantum computing. 
Firstly, it is common for data analysis methods to be proposed that rely upon future technology that is decades in the making due to the challenges involved in the instrumentation of gravitational wave detections.
%\textcolor{red}{Sarah: I didn't understand what you were getting at with this second point, could you rephrase please? The fact that we need repeated analysis seems to suggest that quantum computers would also require reloading of data, which would not be good} 
Second, the computational expense that gravitational wave data analysis often faces is not due to the quantity of data but in the size of the solution space, as opposed to \textit{big data} problems where techniques such as dimensionality reduction and data compression are required.
Although quantum algorithms with super-polynomial speed-ups and potential applications to big data problems have been proposed~\cite{PhysRevLett.103.150502,lloyd2014quantum}, we must be careful to account for the cost of loading classical data into quantum states~\cite{aaronson2015read}. In the worst case, this can be sufficient to negate any quantum advantage. Gravitational wave data analysis problems are not of this flavour.
Lastly, gravitational wave signals are well modelled through general relativity, allowing their waveforms to be efficiently prepared in quantum states.
It is this final aspect that is the focus of this paper, where we demonstrate how analytical gravitational waveforms can be encoded into quantum states. 

There are different methods for encoding information into quantum states. 
Digital encoding involves storing in the computational basis of a string of qubits, just as information is stored on bit strings in the classical case.
A denser format is angular encoding where each qubit can store up to two real values as rotation angles about the Bloch sphere~\cite{larose2020robust}.
The densest form of information representation is \textit{amplitude encoding}, where the components of a classical vector are stored as the amplitudes of a superposition of quantum states, the number of which grows exponentially with the number of qubits.
Amplitude encoded data is an essential assumption in a multitude of quantum machine learning algorithms~\cite{kerenidis2016quantum,wan2017quantum,romero2017quantum,farhi2018classification,schuld2020circuit,bondarenko2020quantum,wang2021quantum,zhang2021generic}.
The efficient amplitude encoding of vectors is also a required subroutine of quantum algorithms for solving linear systems of equations~\cite{PhysRevLett.103.150502,clader2013preconditioned,childs2017quantum}.
While amplitude encoding is the most space efficient, the computational cost of preparing an arbitrary state scales exponentially with the number of qubits~\cite{mottonen2004transformation}, making it extremely inefficient to perform~\cite{aaronson2015read}.
It has however been demonstrated that amplitude encoding can be performed efficiently to prepare specific states~\cite{grover2002creating,zhao2019state}.

In this work, we demonstrate by explicit construction that quantum states that amplitude encode analytic gravitational waveforms for inspiralling binary systems can be prepared efficiently.
This is a crucial first step to study space-efficient quantum algorithms for gravitational wave data analysis, which may be implementable on small, near-term quantum devices. We anticipate that our state preparation algorithm may find application as a subroutine to load templates to e.g. train a variational quantum circuit to distinguish signal from noise~\cite{cerezo2021variational}, or in the longer term to construct an oracle for Grover's search using amplitude encoded data~\cite{gao2022quantum}, or for more sophisticated machine learning techniques~\cite{PhysRevLett.103.150502}.

We outline the proposed state preparation routine in Sec.~\ref{sec:encode}, where we also introduce the necessary background from the literature on the subroutines used in the procedure.
The state preparation routine is demonstrated with an example on a quantum simulator in Sec.~\ref{sec:GWI}.
We relate our state preparation routine to the challenge of analysing binary neutron star inspiral data faced during the third generation detector in Sec.~\ref{sec:ET}.
The conclusion is provided in Sec.~\ref{sec:conc}.

\section{State preparation algorithms}\label{sec:encode}

Our central goal in this paper is to introduce methods for loading gravitational wave templates into a quantum register, using amplitude encoding. To achieve this we make use of the fact that analytical expressions that closely approximate the waveforms of interest may be derived from general relativity. In this section we begin by outlining our approach to state preparation, which first prepares the amplitude and then the phase of the desired coefficients. For each of these steps we draw on existing techniques, and the remainder of the section is then dedicated to introducing the background needed for our implementation.

\subsection{Overview of state preparation procedure}
An arbitrary complex vector $\Vec{h}$ of length $N$ has components $\{\tilde{A}(0)e^{i\Psi(0)},\dots,\tilde{A}(N-1)e^{i\Psi(N-1)}\}$ where $\tilde{A}(i)$ and $\Psi(i)$ are real numbers. These components may be represented as amplitudes of a quantum state $\ket{h}$ given by: 
\begin{equation}\label{equ:targ}
    \ket{h} = \frac{1}{\|\tilde{A}\|}\sum^{2^n - 1}_{j=0}\tilde{A}(j)e^{i\Psi(j)}\ket{j},
\end{equation}
where $\ket{j}$ are computational basis states of ${n=\lceil\log_{2}N\rceil}$ qubits and $\|\cdot\|$ represents the norm of the vector. The state $\ket{h}$ is then said to be an amplitude encoding of the normalised vector $\Vec{h}$ onto the $n$ qubits.
In general preparing an arbitrary state of this form requires exponential resources~\cite{iten2016quantum}, however if these coefficients are given by functions with analytical expressions, there exist more efficient procedures, which we exploit in this work.

To perform the amplitude encoding of Eq.~\ref{equ:targ}, our strategy will be to divide the task into two steps. In the first, we construct an operator $\hat{U}_A$, the role of which is to encode the real amplitudes $\tilde{A}(j)$ such that:
\begin{align}
    \ket{F} &= \hat{U}_{A}\ket{0}^{\otimes n} \nonumber \\
    &= \frac{1}{\|\tilde{A}\|}\sum^{2^{n}-1}_{j=0}\tilde{A}(j)\ket{j}.\label{equ:amptarg}
\end{align}
We propose two methods to implement this operator as a quantum circuit: either by using the Grover-Rudolph algorithm~\cite{grover2002creating}, or by using a trained parameterised quantum circuit~\cite{benedetti2019parameterized}, each described below.

In the second step, we construct an operator $\hat{U}_\Psi$, the role of which is to prepare the phases $\Psi(j)$. Thus, the action on computational basis states $\ket{j}$ is such that:
\begin{equation}\label{equ:phaseprep}
    \hat{U}_{\Psi} \ket{j} = e^{i\Psi(j)}\ket{j}.
\end{equation}
We show below by explicit construction that if the phase is given analytically as a function of $j$ this operator may be implemented efficiently as a quantum circuit. Our procedure is straight-forward and closely related to methods used in the quantum Fourier transform~\cite{barnett2009quantum}.

\begin{figure}[!ht]
	\centering
	\includegraphics[width=0.45\textwidth]{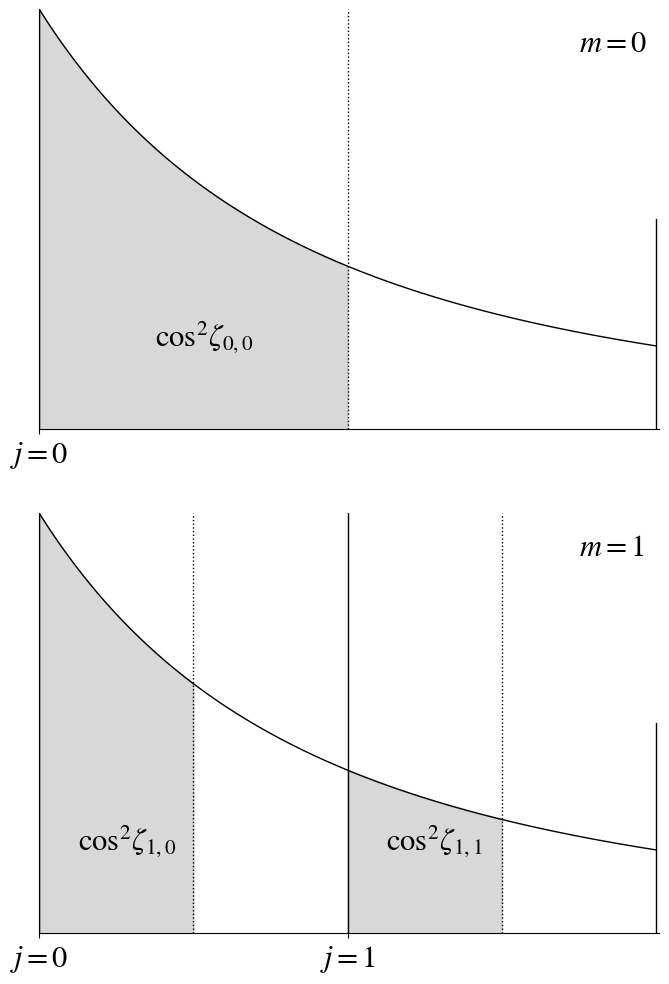}
	\caption{Illustration of the Grover-Rudolph algorithm applied to two qubits to encode $p(f)\propto f^{-7/3}$ (shown as the solid curved line) into the quantum state amplitudes. Initially no qubits are encoded and therefore $m=0$ and there is only one bin $j=0$. The domain is then divided into two and the ratio of the left-most shaded region under the distribution to the whole domain is calculated as $\cos^{2}\zeta_{0,0}$. The $m=0$ qubit is then put into a superposition conditioned on the corresponding $\zeta_{0,0}$ to produce $m=1$. The process is repeated for $m=1$ where $j=0,1$ and each region is split proportionally by $\cos^{2}\zeta_{1,0}$ and $\cos^{2}\zeta_{1,1}$ respectively to acquire $m=2$. Note that a continuous distribution is plotted for illustrative purposes; the discretised probability distribution is considered for formulating $\zeta_{m,j}$ in Eq.~\ref{equ:GR}.}
	\label{fig:GRalg}
\end{figure}

\subsection{Grover-Rudolph algorithm}\label{sec:groverrud}

The Grover-Rudolph algorithm~\cite{grover2002creating} is one way to implement the operator $\hat{U}_A$. This algorithm can efficiently prepare a target quantum state $\ket{\psi}$ of the form:
\begin{equation}\label{equ:GRtarg}
\ket{\psi}=\sum^{2^{n}-1}_{j=0}\sqrt{p(j)}\ket{j},
\end{equation}
given probability mass function $p$. 

The Grover-Rudolph algorithm is an iterative process applied to each ${m=0,\dots,n-1}$ qubits, where each step involves dividing the target distribution into ${j=0,\dots,2^{m}-1}$ bins.
For the $m$th step, each $j$th bin is divided into two further bins of equal width, and the corresponding probability amplitude divided between the two new bins is calculated and stored in an ancillary register.
This is illustrated in Fig.~\ref{fig:GRalg}.
The fraction of the probability residing in the leftmost half of the $j$th bin is equal to:
\begin{equation}\label{equ:GR}
    \cos^{2}\zeta_{m,j} = \frac{\sum_{i=j2^{n-m}}^{(j+1/2)2^{n-m}}p(i)}{\sum_{l=j2^{n-m}}^{(j+1)2^{n-m}}p(i)}. %\arccos\left(\sqrt{\frac{\Delta f2^{1+n-m}(1-(2j)^{-4/3})}{(2(j+1))^{-4/3}-(2j)^{-4/3}}}\right).
\end{equation}
Consider the operation $\hat{Q}^{(m)}_{\zeta}$ that calculates the angle $\zeta_{m,j}$, storing this in the computational basis of an ancillary register $\ket{0}_{\text{a}}$:
\begin{equation}\label{equ:Pzet}
    \hat{Q}_{\zeta}^{(m)}\left(\sum_{j=0}^{2^{m}-1}\sqrt{p^{(m)}_j}\ket{j}\right)\ket{0}_{\text{a}} = \sum_{j=0}^{2^{m}-1}\sqrt{p^{(m)}_j}\ket{j}\ket{\zeta_{m,j}}_{\text{a}},
\end{equation}
where $p^{m}_{j}$ denotes the probability mass function $p$ integrated across the $j$th bin given $m$ qubits. 
Controlled $Y$ rotations $\hat{R}_{Cy}^{(m+1)}$ are then applied onto qubit $m+1$ such that:
\begin{align}\label{equ:Pdef}
            & \hat{R}_{Cy}^{(m+1)}\sqrt{p^{(m)}_j}\ket{j}\ket{\zeta_{m,j}}_{\text{a}}\ket{0}_{m+1} \nonumber\\ & = \sqrt{p^{(m)}_j}\ket{j}\ket{\zeta_{m,j}}_{\text{a}}(\cos\zeta_{m,j}\ket{0}_{m+1}+\sin\zeta_{m,j}\ket{1}_{m+1}) \nonumber\\
       & =\sqrt{p^{(m+1)}_j}\ket{j}\ket{\zeta_{m,j}}_{\text{a}}.
\end{align}
The ancilla register is then cleared by uncomputing $\zeta_{m,j}$ using the inverse operation $\hat{Q}^{(m)}_{\zeta}$ of Eq.~\ref{equ:Pzet}.
The operations of Eq.~\ref{equ:Pzet} and Eq.~\ref{equ:Pdef} can be repeated for $m$ incremented by one, until ${m=n-1}$, and the state of Eq.~\ref{equ:GRtarg} is produced.

While this routine allows for efficient state preparation of the $n$ qubit amplitude states, for practical use we will be interested not just in the scaling of the cost with $n$, but also the absolute number of gates used. The computational cost for low $m$ iterations may induce overhead costs that are much greater than simply using a general state preparation routine for loading arbitrary states~\cite{iten2016quantum}.
Therefore we can define a critical value $m_a$ where for $m<m_a$ a routine for arbitrary amplitude state preparation is applied instead that we can denote $R_{y}^{(m<m_a)}$.
This routine can be further parameterised to reduce computational cost by defining a maximum iteration $m_b$.
As $\lim_{m \to \infty} \zeta_{m,j}=\pi/4$ for all $j$ in Eq.~\ref{equ:GR} (assuming continuity of the probability mass function $p$), then replacing the operation described in Eq.~\ref{equ:Pdef} by applying Hadamard gates to the qubits for operations ${m>m_b}$ reduces the gate cost at the expense of the accuracy of the final amplitude states.
The overall encoding of the frequency amplitude is performed by the time-ordered product
\begin{equation}\label{equ:UA}
    \hat{U}_A = \hat{H}^{(m>m_b)}\left(\prod^{m_b-2}_{m=m_a}\hat{Q}_{\zeta}^{\dag(m)}\hat{R}_{Cy}^{(m+1)}\hat{Q}_{\zeta}^{(m)}\right)\hat{R}_{y}^{(m<m_a)}
\end{equation}
where $\hat{H}^{(m>m_b)}$ denotes Hadamard gates applied onto qubits corresponding to $m>m_b$.

\subsection{Quantum generative modelling}\label{sec:QGAN}

Another approach to preparing a state of the form of Eq.~\ref{equ:GRtarg} is to train a parameterized quantum circuit $\hat{U}(\phi)$ using a hybrid quantum-classical generative machine learning model given parameters $\phi$.
This is done by measuring an ensemble of outputs from the parameterized quantum circuit and utilizing a classical loss function to compare the samples from the measured distribution $q_{\phi}$ to those from the target distribution $p$ which can be classically generated.
This requires the parameterized quantum circuit to be chosen so that it is constrained to only result in real amplitudes. We choose a parameterized circuit of $L$ repeating layers, defined by:
\begin{align}\label{equ:PQC}
\hat{U}(\phi) =& \prod_{j=0}^{n-1}\hat{R}_{y}^{(j)}(\phi_{L,j}) \nonumber \\ 
& \prod_{i=0}^{L-1}\left(\prod_{k=0}^{n-2}\hat{X}^{(k,k+1)}_{C}\prod_{j=0}^{n-1}\hat{R}_{y}^{(j)}(\phi_{i,j})\right)\hat{H}^{\otimes n},
\end{align}
where parameter $\phi_{i,j}$ is the parameter for layer $i$ and qubit $j$, $\hat{R}^{(j)}_{y}$ is a $Y$ rotation on qubit $j$ by the given angle, and $\hat{X}^{(k,k+1)}_{C}$ is an $X$ gate on qubit $k+1$ controlled on qubit $k$. 

We explore the use of a hybrid quantum-classical \textit{generative adversarial network} to train this parameterised quantum circuit~\cite{benedetti2019parameterized}.
Generative adversarial networks consider two competing networks: one, called the \textit{discriminator} and dependent on parameters $\omega$, is trained to discern an ensemble of samples from $q_{\phi}$ from those taken from $p$. The other, called the \textit{generator} and dependent on parameters $\phi$, is trained to produce samples $q_{\phi}$ to trick the discriminator into falsely labeling them as being products from $p$.
As both networks have competing objectives, the training of both simultaneously leads to ${q_{\phi}\approx p}$ as Nash equilibrium is reached~\cite{goodfellow2020generative}.
Quantum generative adversarial networks use a quantum parameterized circuit in place of the generator as discussed in Ref.~\cite{zoufal2019quantum}.
The discriminator given parameters $\omega$ outputs $D_{\omega}$, where ${0<D_{\omega}<1/2}$ indicates the samples were drawn from $q_{\theta}$, while ${1/2<D_{\omega}<1}$ indicates the samples were drawn from $p$.
The training involves minimizing the generator loss function given $S$ samples ${\{x_1,\dots,x_{S}\}\sim q_{\phi}}$:
\begin{equation}\label{equ:genloss}
L_{G} = -\frac{1}{S}\sum_{i=1}^{S}\log D_{\omega}(x_i),
\end{equation}
and maximizing the discriminator loss function given $S$ samples ${\{x_1',\dots,x_S'\}\sim p}$:
\begin{equation}\label{equ:discloss}
L_{D} = \frac{1}{S}\sum_{i=1}^{S}[\log D_{\omega}(x_i') + \log(1-D_{\omega}(x_i))].
\end{equation}

\subsection{Phase preparation}\label{sec:phaseprep}

To encode the phase information we wish to construct an operator $\hat{U}_{\Psi}$, as defined in Eq.~\ref{equ:phaseprep}. Given an analytic expression for $\Psi(j)$, which may be computed efficiently classically, we begin by evaluating $\Psi^\prime(j) = \Psi(j)/2 \pi$ and storing it as a binary string in the computational basis of an ancilla register. We define the operator implementing this as $\hat{Q}_\Psi$:
\begin{equation}\label{equ:Pphase}
    \hat{Q}_{\Psi}\ket{j}\ket{0}_{\text{a}} = \ket{j}\ket{\Psi^{\prime}(j)}_{\text{a}}.
\end{equation}
As this operation can be computed efficiently classically, it follows that $\hat{Q}_\Psi$ may be implemented efficiently as a quantum circuit.

The frequency dependent phase can then be readily produced by $Z$ rotations $\hat{R}_{z}$ applied to the qubits of the ancillary register. Specifically, any number $x$ stored in $n$-bits, $p$ of which are precision qubits, has binary representation:
\[x = \sum^{n-1}_{i=0}x_{i}2^{i-p}.\]
This may be stored in the computational basis of an $n$-qubit register as the state
\[
\ket{x} = \ket{x_{n-1}} \otimes \ket{x_{n-2}} \ldots \otimes \ket{x_0}
\]
Applying single qubit $Z$ rotations to the precision qubits, where $\hat{R}^{(j)}(2^{j-p+1}\pi)$ represents a rotation by $2^{j-p+1}\pi$ applied to the $j$-th qubit gives:
\begin{align}
\prod e^{2^{j-p+1}\pi x_j} \ket{x_{n-1}} \otimes \ket{x_{n-2}} \ldots \ket{x_0} \nonumber\\
= e^{2 \pi \sum_{j=0}^{p-1} x_j 2^{j-p}} \ket{x} = e^{2 \pi x} \ket{x}
\end{align}.
Thus we need only apply $p_a$ single qubit gates (where $p_a$ is the number of precision qubits used in the ancilliary register) to prepare:
\begin{equation}\label{equ:PCRZ}
    \prod^{p_a}_{j=1}\hat{R}_{z}^{(p_a-j)}(2^{1-j}\pi)\ket{\Psi^{\prime}(j)}_{\text{a}} = e^{i\Psi(j)}\ket{\Psi^{\prime}(j)}_{\text{a}}.
\end{equation}

Finally, the ancillary register is cleared by uncomputing the calculation of $\Psi^{\prime}(j)$ with the inverse operation $\hat{Q}^{\dag}_{\Psi}$ giving
\begin{equation}\label{equ:Rz}
    \hat{U}_{\Psi} = \hat{Q}_{\Psi}^{\dag}\prod^{p_a}_{j=1}\hat{R}_{z}^{(p_a-j)}(2^{1-j}\pi)\hat{Q}_{\Psi}.
\end{equation}

\subsection{Linear piece-wise function}\label{sec:LPWF}

\begin{figure*}[!ht]
\centering
%\scalebox{1.}{
\[\Qcircuit @C=.3em @R=.02em @! {
\lstick{\ket{0}_{l}} & \multigate{1}{\text{Label}} & \ctrl{1} & \qw & \ctrl{1} & \ctrl{1} & \qw & \qw \\%\ctrl{1} & \multigate{1}{\text{Label}^{\dag}} & \qw \\
\lstick{\ket{0}_{\text{c}}} & \ghost{\text{Label}} & \gate{X_{1}} & \multigate{2}{\text{Mult}} & \gate{X_{1}^{\dag}} & \gate{X_{0}} & \multigate{1}{\text{Add}} & \qw \\%\gate{X_{0}^{\dag}} & \ghost{\text{Label}^{\dag}} & \qw \\
\lstick{\ket{0}_o} & \qw & \qw & \ghost{\text{Mult}} & \qw & \qw & \ghost{\text{Add}} & \rstick{\ket{f(x)}_o} \qw \\
\lstick{\ket{x}_x} & \ctrl{-2} & \qw & \ghost{\text{Mult}} & \qw & \qw & \qw & \qw\\%& \ctrl{-2} & \qw %\gategroup{1}{3}{4}{8}{2.5em}{--}
}\]
%}
\caption{Quantum circuit of the piece-wise linear function operation $\hat{Q}_{f}$ approximating $f(x)\approx A_1^{l} x + A_0^{l}$ for values of $x$ within each of the $2^{n_{l}}$ sub-domains of $f(x)$. The label gate correlates the sub-domains to the label register $\ket{l}_l$ before loading in the coefficients $A_{1}^{l}$ for each sub-domain into the coefficient register using controlled operations $X_{1}$. The input register $\ket{x}_x$ is then multiplied by the coefficient register with the arithmetic multiplication operation and the result is saved in the output register. The coefficient register is cleared before the zeroth order coefficients are loaded with operation $X_0$ before they are added to the output register with an arithmetic addition operation.}
\label{fig:Pgate}
\end{figure*}
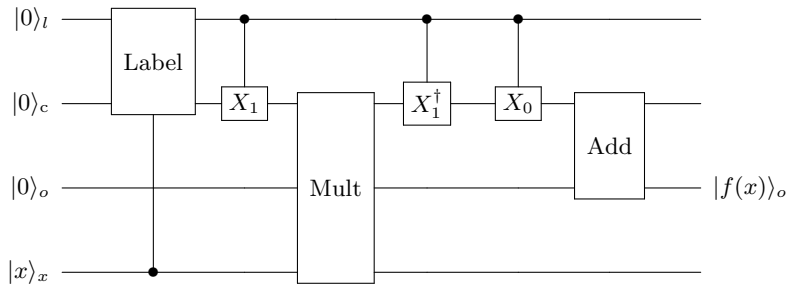

For both Eq.~\ref{equ:Pzet} of the Grover-Rudolph algorithm and Eq.~\ref{equ:Pphase} of the phase preparation, we require a quantum circuit that is able to efficiently input and evaluate functions $f(x)$ for a given $x$.
Any function that is computable efficiently classically may be directly implemented as a quantum circuit, performing the function evaluation in the computational basis and writing the result in an ancilla register:
\begin{equation*}
    \ket{x} \ket{0} \rightarrow \ket{x} \ket{f(x)}\end{equation*}
One way to implement function evaluation in practice is by a piece-wise function approximation, described in Ref.~\cite{haner2018optimizing}, which has previously been proposed as a way to perform the Grover-Rudolph algorithm efficiently in Ref.~\cite{wang2021efficient}.
This involves dividing the domain into $2^{n_{l}}$ sub-domains and approximating the function in each sub-domain by a polynomial of a given order.
The Remez Algorithm~\cite{remez1934determination} is applied to determine the polynomial coefficients to the given order that minimizes the $L_{\infty}$ error across the given domain.
This classical pre-processing step is performed with a cost of $O(2^{n_{l}})$.
The domains of the function are then correlated with a label register $\ket{0}^{\otimes n_{l}}_{l}$ using the label gate described in Ref.~\cite{haner2018optimizing}.
Here we consider the simplest case where the function $f(x)$ is approximated to be linear within the given sub-domain over $x$, requiring the zeroth and first order polynomial coefficients $A_{0}^{l}$ and $A_{1}^{l}$ such that $f(x)\approx A_1^{l} x + A_0^{l}$ for $x$ in the given sub-domain.
The $x$ argument is given to the input register $\ket{x}_x^{\otimes n_x}$ of size $n_x$ and the coefficients for each sub-domain are loaded into the coefficient register $\ket{0}^{\otimes n_{c}}_{c}$, where $n_c$ qubits are used.
This requires the output register in which the outcome $f(x)$ is stored in $n_{\text{o}}$ qubits.
Generally, when approximating a linear piece-wise function, we introduce the operation $\hat{Q}_f$, performing the operations:
\begin{align*} 
\ket{x}_x\ket{0}_o\ket{A_1^{l}}_c\ket{l}_l &\xrightarrow{\text{Mult}} \ket{x}_x\ket{A_1^{l} x}_o\ket{A_1^{l}}_c\ket{l}_l \\
&\xrightarrow{\text{Load}} \ket{x}_x\ket{A_1^{l} x}_o\ket{A_0^{l}}_c\ket{l}_l \\
&\xrightarrow{\text{Add}} \ket{x}_x\ket{A_1^{l} x + A_0^{l}}_o\ket{A_0^{l}}_c\ket{l}_l.
\end{align*}
The quantum circuit used to perform this action is shown in Fig.~\ref{fig:Pgate}, where the zeroth and first order coefficients are loaded and unloaded using gates $X_{0,1}$ and $X_{0,1}^{\dag}$ respectively.
The gate and space cost of applying the piece-wise linear function depends on the choice of adder and multiplier routine.
The coefficients of a given domain are loaded and unloaded into the coefficient register using controlled $X$ gates conditioned on the label register.

\section{Encoding gravitational wave inspirals}\label{sec:GWI}

In this section we detail how the techniques outlined in the previous section are applied to encode the waveforms of gravitational wave binary system inspirals.
An analytical expression of the frequency dependent amplitude of an inspiral waveform is obtained using the stationary-phase approximation (see Appendix~\ref{sec:SPA}), resulting in the expression~\cite{poisson1995gravitational}:
\begin{equation}\label{equ:Af}
    \tilde{A}_{N}(f) = \frac{Q\mathcal{M}^{5/6}}{D}f^{-7/6}.
\end{equation}
where $Q$ is dependent on the geometry of the detector and source system, $\mathcal{M}$ is the chirp mass, and $D$ is the luminosity distance to the source.
Similarly, the frequency dependent phase to a second post-Newtonian order waveform is of the form:
\begin{widetext}
\begin{multline}\label{equ:Psif2N}
    \Psi_{\text{2PN}}(f) = 2\pi ft_c - \phi_c - \frac{\pi}{4} + \frac{3}{128}(\pi\mathcal{M}f)^{-5/3}\left[1+\frac{20}{9}\left(\frac{743}{336} + \frac{11}{4}\eta\right)(\pi Mf)^{2/3} - 4(4\pi - \beta)(\pi Mf) \right. \\ \left. + 10\left(\frac{3058673}{1016064} + \frac{5429}{1008}\eta + \frac{617}{144}\eta^{2} - \sigma\right)(\pi Mf)^{4/3} \right].
\end{multline}
\end{widetext}
where $t_c$ is the time of coalescence, $\phi_c$ is the phase at coalescence, $M$ is the total mass, $\eta$ is the reduced mass, $\sigma$ is the spin-spin parameter, and $\beta$ is the spin-orbit parameter.
For further details on these parameters, we refer the reader to Appendix~\ref{sec:GWCBC}.

Consider encoding the frequency dependent waveform over a frequency range with lower frequency cut-off $f_{\text{min}}$ and upper frequency $f_{\text{max}}$, discretised into ${N=2^{n}}$ frequency bins of widths ${\Delta f = (f_{\text{max}}-f_{\text{min}})/2^{n}}$.
This gives the number of integer qubits to be ${n_{\text{int}} = \lceil \log_2 (2^{n}\Delta f)\rceil}$ and therefore the number of precision bits as ${p = \lceil \log_{2}T\rceil}$, where $T$ is the waveform's temporal duration ${T=\Delta f^{-1}}$.
The $j$th computational basis state has a probability amplitude equal to the integrated real classical frequency amplitude within the frequency range ${f_{\text{min}}+[j\Delta f,(j+1)\Delta f)}$.

While the Newtonian waveform amplitude depends on the detector response, chirp-mass and source distance, only the frequency dependence is required in the encoding due to the normalisation, and therefore these terms are ignored such that ${\tilde{A}(f)\propto f^{-7/6}}$.

To demonstrate the state-preparation routine of Sec.~\ref{sec:encode}, we prepare a waveform with a Newtonian amplitude of Eq.~\ref{equ:Af} and a second post-Newtonian phase of Eq.~\ref{equ:Psif2N}, given a spinless (${\beta=\sigma=0}$), near equal-mass system of ${m_1=35\,M_{\odot}}$ and ${m_2=30\,M_{\odot}}$ using IBM's quantum simulator and qiskit software~\cite{Qiskit}.
This simulation is carried out on Python code that is publicly available on Github~\cite{hayesgithub}.
The simulation is run on a commercial computer with 7.9 gigabytes of available memory, allowing for simulations of up to 28 qubits when allowing complex numbers to be stored on 16 bytes.
This waveform is sampled within the frequency interval ${f_{\text{min}}=40}\,$Hz and ${f_{\text{max}}=168}\,$Hz with ${\Delta f=2}\,$Hz, requiring ${n=6}$ qubits.
A value of $\Delta T = 0.02\,$s is chosen. 

After the simulation, the output waveform state $\ket{\psi_{\text{out}}}$ is compared to the classically computed target waveform state $\ket{\psi_{\text{targ}}}$ and the \textit{fidelity} $\mathcal{F}$ between the two states is calculated, defined as $|\braket{\psi_{\text{out}}}{\psi_{\text{targ}}}|^{2}$, such that ${1-\sqrt{\mathcal{F}}}$ is synonymous with the \textit{mismatch} associated between templates as defined in \cite{owen1996search} where the noise is flat across all frequencies.

\begin{figure*}[t!]
    \centering
    \begin{subfigure}[t]{0.5\textwidth}
    \includegraphics[width=1\textwidth]{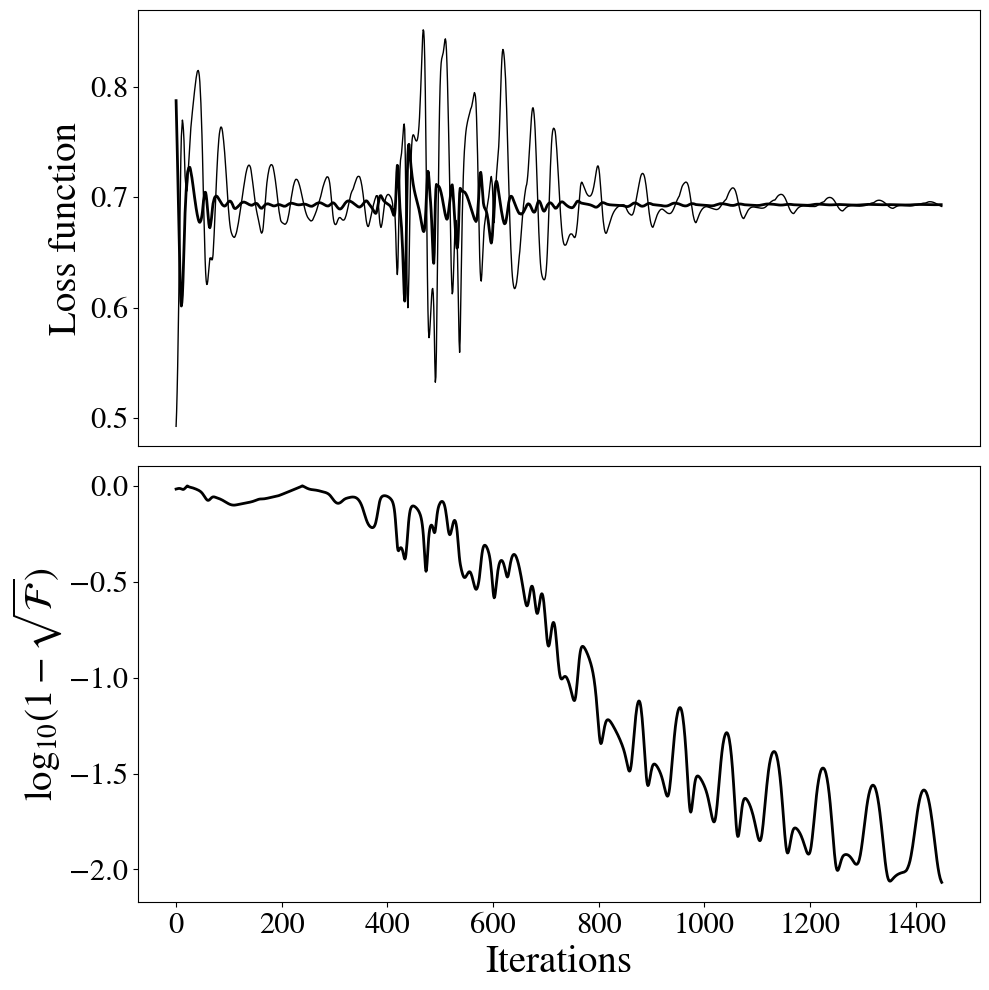}
	\caption{}
	\label{fig:QGANtrain}
    \end{subfigure}%
    ~ 
    \begin{subfigure}[t]{0.5\textwidth}
	\includegraphics[width=1\textwidth]{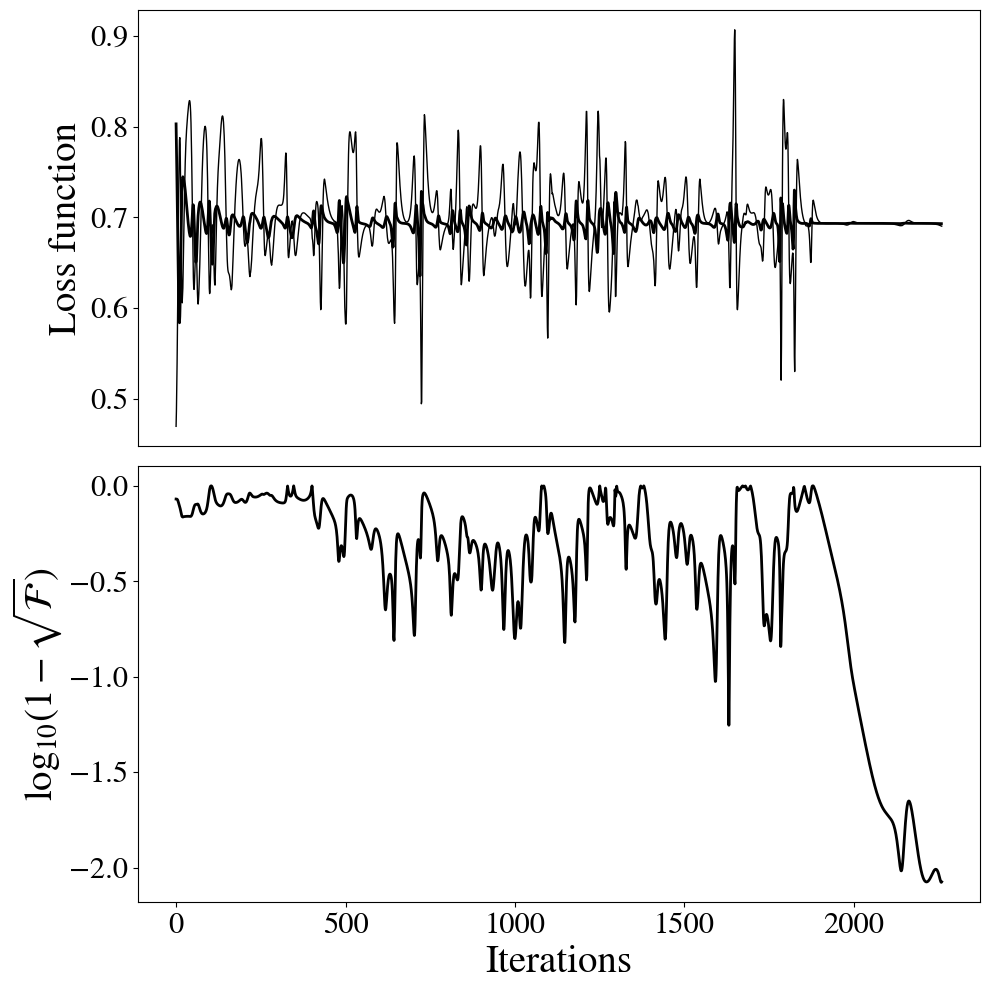}
	\caption{}
	\label{fig:QGANtrainbig}
    \end{subfigure}
    \caption{The top panel shows the loss over training iterations for the quantum generative adversarial network configuration for both the generator shown as the black solid line as defined in Eq.~\ref{equ:genloss}, and the discriminator in Eq.~\ref{equ:discloss} as the grey solid line. The dialectic between both networks is represented by the loss, as they exhibit opposing gradients about an equilibrium of ${\sim}0.7$. The bottom panel shows the resulting generated quantum state mismatch with the target state amplitude of $\tilde{A}(f)\propto f^{-7/6}$ over iterations. $(a)$ shows the training of a parameterized quantum circuit of $L=12$ with $78$ trainable parameters, which reaches a minimum mismatch of $8.57{\times}10^{-3}$. $(b)$ shows the training of a parameterized quantum circuit of $L=20$ with $126$ trainable parameters, which reaches a minimum mismatch of $8.36{\times}10^{-3}$.}
    \label{fig:QGANtrainboth}
\end{figure*}

\begin{figure*}[t!]
    \centering
    \begin{subfigure}[t]{0.49\textwidth}
	\includegraphics[width=1\textwidth]{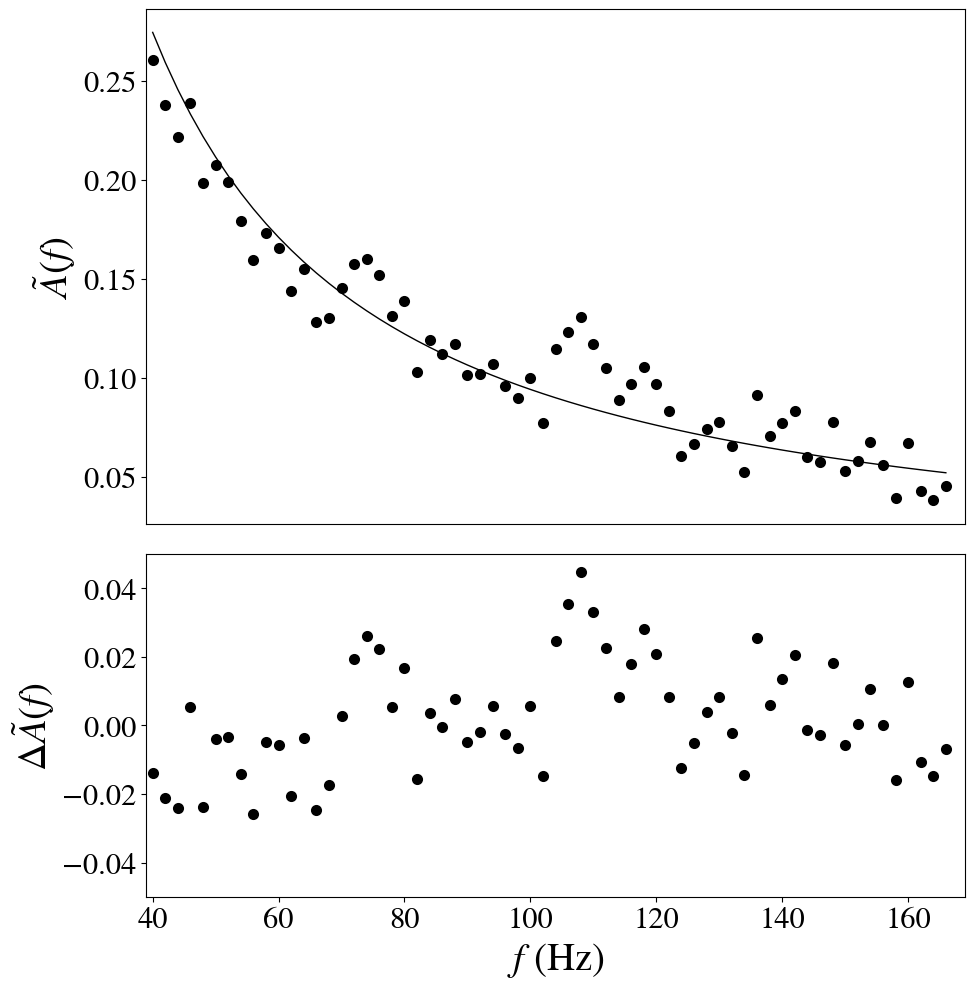}
	\caption{}
	\label{fig:QGANAf}
    \end{subfigure}
    ~
    \begin{subfigure}[t]{0.49\textwidth}
    \includegraphics[width=1\textwidth]{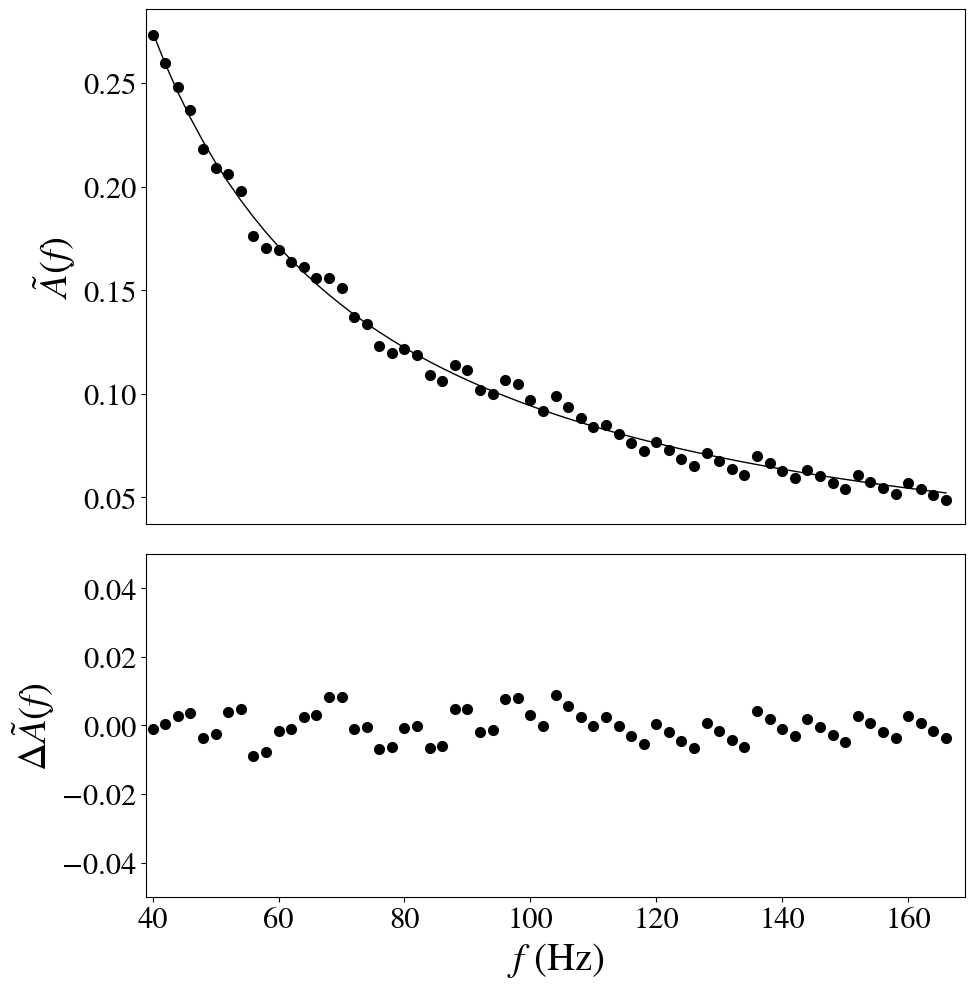}
	\caption{}
	\label{fig:Af}
    \end{subfigure}%
    \caption{A scatter of simulated frequency state amplitude compared to the target state amplitudes of $\tilde{A}(f)\propto f^{-7/6}$ shown as the solid black line in the top panel. The bottom panel shows the relative difference between the output and the target frequency states. $(a)$ shows the resulting amplitudes from simulating the parameterized quantum circuit of Eq.~\ref{equ:PQC} after the training illustrated in Fig.~\ref{fig:QGANtrain}. The resulting state fidelity is $0.983$ with a mismatch of $8.6{\times}10^{-3}$. Discrepancies between the simulated and target state are mainly due to the limited size of the parameterized quantum circuit in this case. The number of controlled NOT gates used in the simulation is $100$. $(b)$ shows the resulting amplitudes from simulating the circuit described in Fig.~\ref{fig:GRcirc} resulting in a state fidelity of $0.999$ and mismatch of $4.1{\times}10^{-4}$. Discrepancies between the two are caused by the limited number of qubits available for the arithmetic operations involved in the Grover-Rudolph algorithm when invoking the piece-wise linear function operation as well as the omission of second-order terms. The number of controlled NOT gates used is $23,796$.}
\end{figure*}

\subsection{Amplitude preparation}

\begin{figure}[ht]
\centering
%\scalebox{1.}{
\[\Qcircuit @C=.0em @R=.0em @! {
\lstick{\ket{0}_{l}} & \qw & \multigate{7}{Q_{\zeta}^{(4)}} & \qw & \multigate{7}{Q_{\zeta}^{\dag (4)}} & \qw \\% \multigate{8}{Q_{\zeta}^{(5)}} & \qw & \multigate{8}{Q_{\zeta}^{\dag (5)}} \barrier[-2.0em]{10} & \multigate{9}{Q_{\zeta}^{(6)}} & \qw & \multigate{9}{Q_{\zeta}^{\dag (6)}} & \qw \\
\lstick{\ket{0}_{c}} & \qw & \ghost{Q_{\zeta}^{(4)}} & \qw & \ghost{Q_{\zeta}^{\dag (4)}} & \qw \\% \ghost{Q_{\zeta}^{(5)}} & \qw & \ghost{Q_{\zeta}^{\dag (5)}} \barrier[-2.0em]{9} & \ghost{Q_{\zeta}^{(6)}} & \qw & \ghost{Q_{\zeta}^{\dag (6)}} & \qw \\
\lstick{\ket{0}_{\text{a}}} & \qw & \ghost{Q_{\zeta}^{(4)}} & \ctrl{6} & \ghost{Q_{\zeta}^{\dag (4)}} & \qw \\% \ghost{Q_{\zeta}^{(5)}} & \ctrl{7} & \ghost{Q_{\zeta}^{\dag (5)}} & \ghost{Q_{\zeta}^{(6)}} & \ctrl{8} & \ghost{Q_{\zeta}^{\dag (6)}} & \qw \\
\ustick{0} & \multigate{4}{R_{Y}^{(m<5)}} & \ghost{Q_{\zeta}^{(4}} & \qw & \ghost{Q_{\zeta}^{\dag (4)}} & \qw \\% \ghost{Q_{\zeta}^{(5)}} & \qw & \ghost{Q_{\zeta}^{\dag (5)}} & \ghost{Q_{\zeta}^{(6)}} & \qw & \ghost{Q_{\zeta}^{\dag (6)}} & \qw \\
\ustick{1} & \ghost{R_{Y}^{(m<5)}} & \ghost{Q_{\zeta}^{(4)}} & \qw & \ghost{Q_{\zeta}^{\dag (4)}} & \qw \\% \ghost{Q_{\zeta}^{(5)}} & \qw & \ghost{Q_{\zeta}^{\dag (5)}} & \ghost{Q_{\zeta}^{(6)}} & \qw & \ghost{Q_{\zeta}^{\dag (6)}} & \qw \\
\ustick{2} & \ghost{R_{Y}^{(m<5)}} & \ghost{Q_{\zeta}^{(4)}} & \qw & \ghost{Q_{\zeta}^{\dag (4)}} & \qw \\% \ghost{Q_{\zeta}^{(5)}} & \qw & \ghost{Q_{\zeta}^{\dag (5)}} & \ghost{Q_{\zeta}^{(6)}} & \qw & \ghost{Q_{\zeta}^{\dag (6)}} & \qw \\
\ustick{3} & \ghost{R_{Y}^{(m<5)}} & \ghost{Q_{\zeta}^{(4)}} & \qw & \ghost{Q_{\zeta}^{\dag (4)}} & \qw \\% \ghost{Q_{\zeta}^{(5)}} & \qw & \ghost{Q_{\zeta}^{\dag (5)}} & \ghost{Q_{\zeta}^{(6)}} & \qw & \ghost{Q_{\zeta}^{\dag (6)}} & \qw \\
\ustick{4} & \ghost{R_{Y}^{(m<5)}} & \ghost{Q_{\zeta}^{(4)}} & \qw & \ghost{Q_{\zeta}^{\dag (4)}} & \qw \\% \ghost{Q_{\zeta}^{(5)}} & \qw & \ghost{Q_{\zeta}^{\dag (5)}} & \ghost{Q_{\zeta}^{(6)}} & \qw & \ghost{Q_{\zeta}^{\dag (6)}} & \qw \\
\ustick{5} & \qw & \qw & \gate{R_{Y}^{(5)}} & \qw & \qw \\% \ghost{Q_{\zeta}^{(5)}} & \qw & \ghost{Q_{\zeta}^{\dag (5)}} & \ghost{P_{\zeta^{(6)}}} & \qw & \ghost{Q_{\zeta}^{\dag (6)}} & \qw \\
%\ustick{6} & \qw & \qw & \qw & \qw & \qw & \gate{R_{Y}^{(6)}} & \qw & \ghost{Q_{\zeta}^{(6)}} & \qw & \ghost{Q_{\zeta}^{\dag (6)}} & \qw \\
%\ustick{7} & \qw & \qw & \qw & \qw & \qw & \qw & \qw & \qw & \gate{R_{Y}^{(7)}} & \qw & \qw \inputgroupv{4}{11}{1.1em}{15.5em}{\ket{0}_{h}}
}\]
%}
\caption{The circuit of the amplitude preparation step using the Grover-Rudolph algorithm where the steps for $m=0,1,2,3,4$ are performed by controlled rotations of angles $\zeta_{m,j}$ for all $j=0,\dots,2^{m}-1$ denoted by operation $R_{y}^{(m<5)}$, while the operations for $m=5$ are performed by the operations $\hat{Q}^{\dag (m-1)}_{\zeta}\hat{R}_{y}^{(m)}\hat{Q}_{\zeta}^{(m-1)}$.}
\label{fig:GRcirc}
\end{figure}
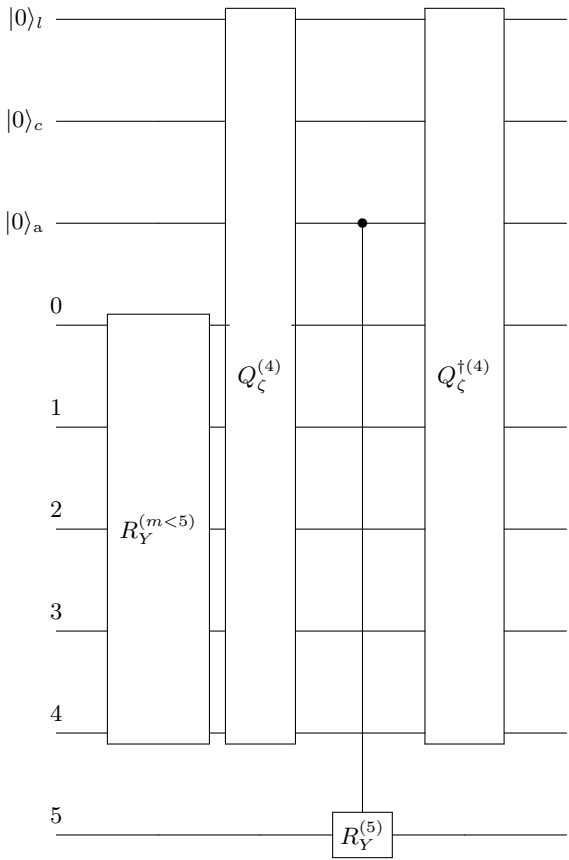

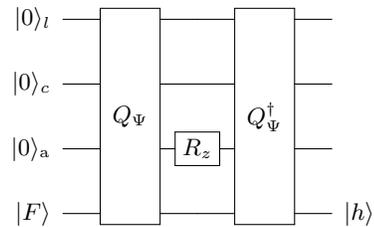
\begin{figure}[!ht]
	\centering
        \[\Qcircuit @C=.3em @R=.2em @! {
        \lstick{\ket{0}_{l}} & \multigate{3}{Q_{\Psi}} & \qw & \multigate{3}{Q_{\Psi}^{\dag}} & \qw  \\
        \lstick{\ket{0}_{c}} & \ghost{Q_{\Psi}} & \qw & \ghost{Q_{\Psi}^{\dag}} & \qw  \\
        \lstick{\ket{0}_{\text{a}}} & \ghost{Q_{\Psi}} & \gate{R_{z}} & \ghost{Q_{\Psi}^{\dag}} & \qw \\
        \lstick{\ket{F}} & \ghost{Q_{\Psi}} & \qw & \ghost{Q_{\Psi}^{\dag}} & \rstick{\ket{h}} \qw
        }\]
	\caption{Quantum circuit for the phase preparation step where the function $\Psi^{\prime}$ is encoded into the computational basis of the ancillary register $\ket{0}_{\text{a}}$ through the operation $\hat{Q}_{\Psi}$, implemented through the linear piece-wise function using coefficient and label registers $\ket{0}_c$ and $\ket{0}_l$. The phase $e^{i \Psi(j)}$ on computational basis state $\ket{j}$ is produced through $Z$ rotations on the precision bits of the binary string stored in the ancillary register. Finally, the ancillary, label and coefficient registers are cleared by the inverse operation $\hat{Q}_{\Psi}^{\dag}$.}
	\label{fig:phaseprep}
\end{figure}

We demonstrate the preparation of the real amplitudes using both the Grover-Rudolph algorithm method and a parameterized quantum circuit trained using a generative adversarial network for this case.

\subsubsection{Quantum generative adversarial network}\label{sec:QGANsim}

First the quantum generative adversarial network described in Sec.~\ref{sec:QGAN} is applied for the case of $n=6$.
This is done by assuming the parameterized quantum circuit of Eq.~\ref{equ:PQC}.
The parameterized circuit is trained using qiskit's quantum generative adversarial network implementation with \textit{PyTorch}~\cite{paszke2019pytorch}.
The classical learning is performed using the Adam optimizer with a learning rate of $0.01$, and first and second momentum parameters of $0.7$ and $0.999$.
The training was run for $1,500$ iterations, taking $10,000$ samples from the generated quantum state and the target distribution to pass to the classical discriminator.
The results of the training are displayed in Fig.~\ref{fig:QGANtrainboth} for a network size of ${L=12}$ and ${L=20}$ in Fig.~\ref{fig:QGANtrain} and Fig.~\ref{fig:QGANtrainbig} respectively.
For both cases, the loss function of both the generator and discriminator are plotted in the top panel (black and grey respectively) over training iterations, and can be seen to oscillate about an equilibrium of ${\sim}0.7$ as the decrease of loss of one network necessarily requires the increase of loss of the other.
The bottom panels display the corresponding mismatch between states generated by the parameterised quantum circuit and the target state at different points of the training. 
The decrease in the mismatch between generated and target state corresponds to the loss functions of both generator and discriminator converging to the equilibrium values.
Both networks converge to a similar mismatch of $8.57\times10^{-3}$ and $8.36\times10^{-3}$ for the case of ${L=12}$ and ${L=20}$ respectively. 
While not plotted, further training proved to result in a mismatch increase for both cases.
The mismatch of Fig.~\ref{fig:QGANtrain} has a more gradual decline than Fig.~\ref{fig:QGANtrainbig}, where the convergence to a desired solution occurs more suddenly and generally after more training iterations.
These two training cases exemplify a general trend found that while increasing the network size can improve the mismatch, the training becomes more sporadic and less stable.
The work required to stabilize this training for larger network sizes is beyond the scope of this paper.

The resulting amplitudes after applying the trained parameterized quantum circuit with $L=20$ is shown in Fig.~\ref{fig:QGANAf} scattered in black and compared to the target state shown as the solid black line, and the relative difference between the two amplitudes scattered in the bottom panel.
The fidelity between the output state and the target state of $0.983$ is achieved, corresponding to a mismatch of $8.36\times10^{-3}$.
The simulation is performed using $100$ controlled NOT gates.

\subsubsection{Grover-Rudolph algorithm}\label{sec:GRsim}

The amplitude preparation step using the Grover-Rudolph algorithm described in Sec.~\ref{sec:groverrud} is applied to obtain the state described in Eq.~\ref{equ:amptarg}.
The circuit performing the operation $\hat{U}_{A}$ is shown in Fig.~\ref{fig:GRcirc}. 
The operation of Eq.~\ref{equ:Pzet} is performed using simple controlled $Y$ rotation and $X$ gates for $m<5$, denoted $\hat{R}_{Y}^{(m<5)}$, and a piece-wise linear function approximation applied to higher values.
The size of the ancilla and coefficient registers is chosen so that $n_{c}=9$ and $n_{\text{o}}=9$, and the domain of the function is divided into $16$ sub-domains using $n_l=4$ qubits where the boundaries of each sub-domain are spread uniformly between the domain of the function.

\begin{figure}[!ht]
\centering
\includegraphics[width=\columnwidth]{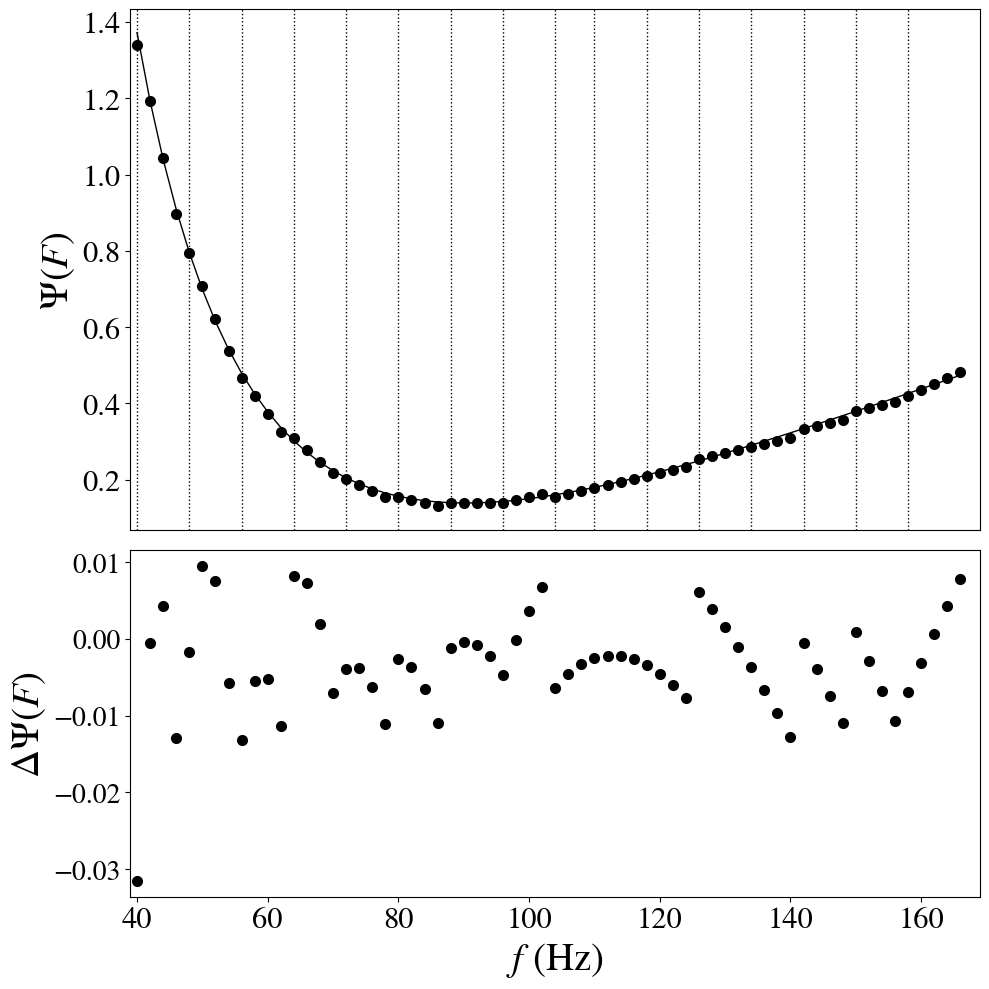}
\caption{The top panel shows a scatter of the simulated values of $\Psi^{\prime}$ stored in the computational basis in the ancillary register for each frequency register state compared to the target function, plotted as the solid line. The boundaries of the piece-wise linear function domains are shown in the dashed vertical lines. The deviations between the values stored in the ancillary register and the target function are scattered in the bottom panel. The simulated states show a maximum deviation from the target function of $|\Delta\Psi|<0.04$. The number of controlled NOT gates used in the simulation is $9,464$.}
\label{fig:Psif}
\end{figure}

\begin{figure*}[t!]
    \centering
    \begin{subfigure}[t]{0.49\textwidth}
	\includegraphics[width=1\textwidth]{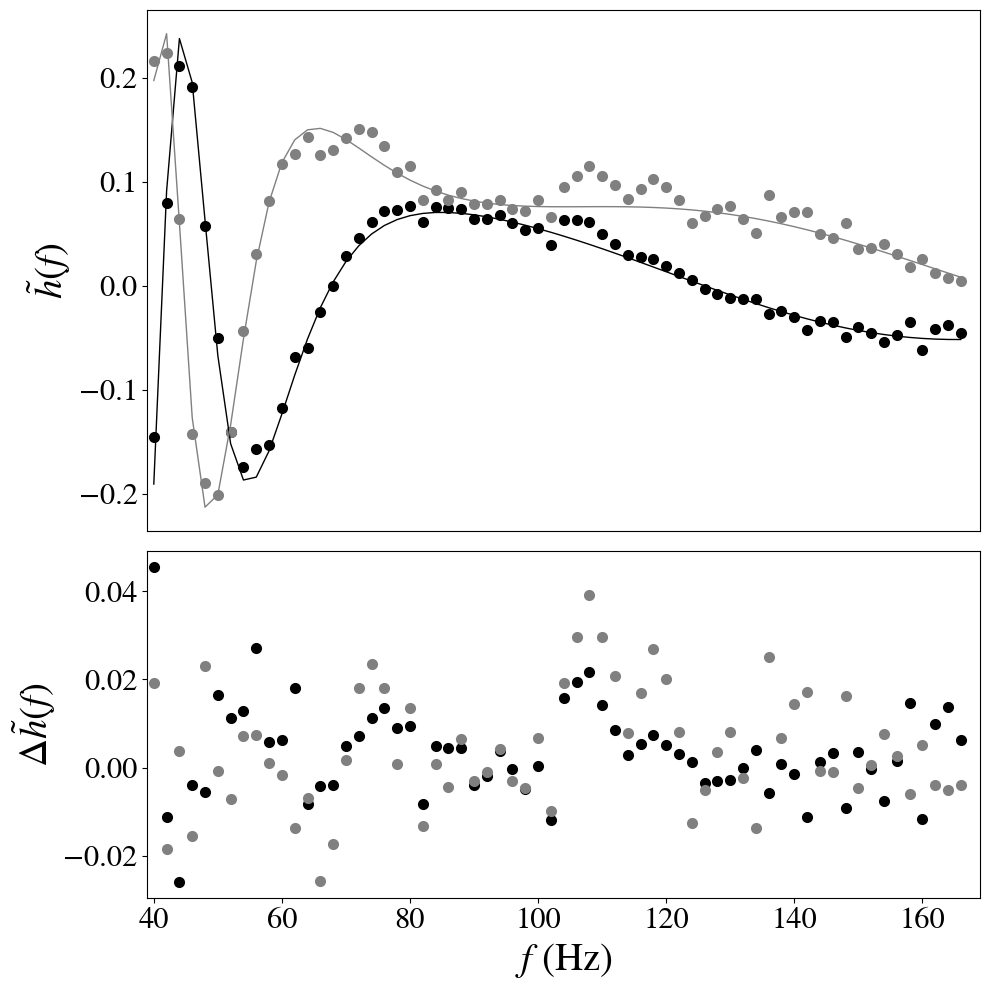}
	\caption{}
	\label{fig:hfQGAN}
    \end{subfigure}
    ~
    \begin{subfigure}[t]{0.49\textwidth}
    \includegraphics[width=1\textwidth]{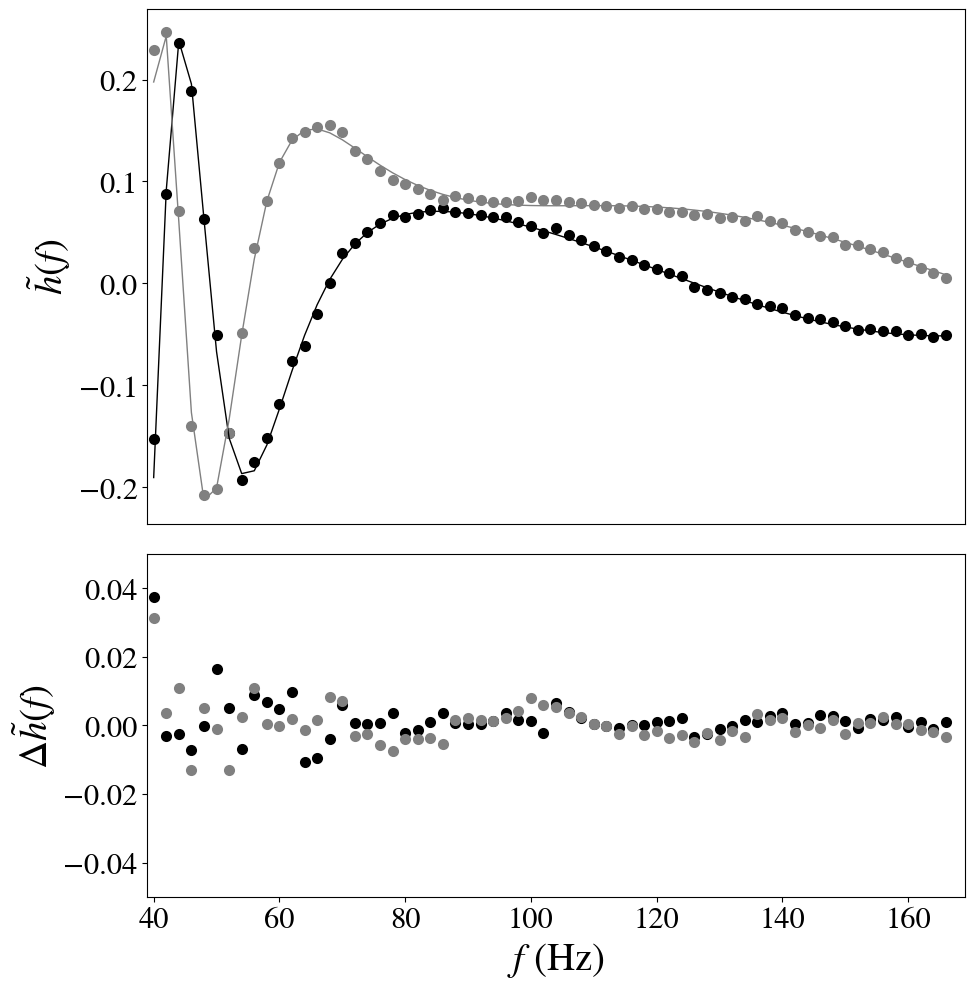}
	\caption{}
	\label{fig:ePsif}
    \end{subfigure}%
    \caption{The top panel shows the simulated real and imaginary amplitudes of the frequency register states, scattered in black and grey respectively. The simulated state amplitudes are compared to the target normalised gravitational waveform, plotted in black and grey solid lines. The deviations between the simulated and the target waveform are scattered in the bottom panel. $(a)$ shows the result after applying the trained parameterized quantum circuit to approximate the amplitude preparation routine as described in Sec.~\ref{sec:QGANsim}. The simulated waveform has a fidelity of $0.979$ and corresponding mismatch of $1.0{\times}10^{-2}$ with the target state. The number of controlled NOT gates used in the simulation is $19,028$. $(b)$ gives the same result but after the phase preparation routine described in Sec.~\ref{sec:phasesim} after performing the Grover-Rudolph amplitude preparation routine described in Sec.~\ref{sec:GRsim} where the simulated waveform has a fidelity of $0.995$ and corresponding mismatch of $2.4\times10^{-3}$ with the target state. The number of controlled NOT gates used is $42,724$.}
\end{figure*}

After the application of the circuit shown in Fig.~\ref{fig:GRcirc}, the amplitude of the output state $\ket{F}$ is plotted in the top panel of Fig.~\ref{fig:Af} as black dots and compared to the target state shown as the solid black line.
The fidelity between the output state and the target state is calculated to be $0.999$ corresponding to a mismatch of $4.1{\times}10^{-4}$, while the relative difference in amplitude between each output and target frequency state is plotted in the bottom panel.
Deviations from the target state are due to the limited number of ancillary qubits to store the rotation angles $\zeta_{m,j}$, load the polynomial coefficients and define the sub-domains of the function, as well as the omission of higher order terms from the linear function approximation.
The simulation is performed using $23,796$ controlled NOT gates.

\subsection{Phase preparation}\label{sec:phasesim}

For the phase preparation step, the operation $\hat{Q}_{\Psi}$ of Eq.~\ref{equ:Pphase} is again applied by the piece-wise linear function approximation described in Sec.~\ref{sec:LPWF}.
The sizes of the coefficient and ancillary registers are set to $n_c=8$ and $n_{\text{a}}=10$ given the piece-wise function coefficients, while the label register size remains unchanged with $n_l=4$.
The resulting joint state across both the frequency and ancillary register corresponds to a superposition in which each term has a single binary string representing $\Psi^{\prime}$ stored in the ancilla registers, correlated to the corresponding frequency state in the frequency register.
The $\Psi^{\prime}$ values for each of the binary strings are scattered in the top panel of Fig.~\ref{fig:Psif} across their corresponding frequency states and compared to the target function $\Psi_{\text{PN}}/2\pi$, plotted as the solid black line.
The resulting joint state amplitudes of Fig.~\ref{fig:Psif} clearly follow a smooth continuous function comparable to the target function.
The bottom panel shows the relative difference between the target function $\Psi_{\text{PN}}/2\pi$ and those stored in the ancillary register, which deviate only such that $|\Delta\Psi|<0.04$.
The boundaries of the piece-wise polynomial function are shown as vertical dotted lines and are uniformly spread across the space.
This step is simulated using $9,464$ controlled NOT gates.

The $\hat{R}_{z}$ rotations of Eq.~\ref{equ:Rz} are applied to the ancilla register that $\Psi^{\prime}$ in the computational basis to produce the frequency dependent phase of Eq.~\ref{equ:PCRZ}.
The ancilla register is then uncomputed of $\Psi^{\prime}$ with the inverse piece-wise linear function, costing an additional $9,464$ controlled NOT gates.
With enough qubits to adequately account for the required precision of the multiplication operation, the uncomputing of $\Psi^{\prime}$ leaves the ancillary and label register in a singular state of $\ket{0}_{l}\ket{0}_{\text{a}}$, which can now be discarded from the circuit.
Fig.~\ref{fig:hfQGAN} and Fig.~\ref{fig:ePsif} depicts the amplitudes of the final simulated states compared to the target state for when the trained parameterised quantum circuit of Fig.~\ref{fig:QGANtrainbig} and the Grover-Rudolph algorithm are used for amplitude preparation steps respectively.
The real and imaginary parts of the target state are plotted as solid lines in their respective colours, with the difference between the output and target state displayed in the bottom panel.
When the Grover-Rudolph algorithm is employed, the resulting state fidelity is $\mathcal{F}=0.995$ with an associated mismatch of $2.4{\times}10^{-3}$ using $42,724$ controlled NOT gates, while when the trained parameterised quantum circuit is used, the state fidelity is $\mathcal{F}=0.979$ with a mismatch of $1.0{\times}10^{-2}$ using $19,028$ controlled NOT gates.

\section{Case study: third generation gravitational wave detectors}\label{sec:ET}

%\begin{figure*}[!ht]
\begin{figure}[!ht]
\centering
\includegraphics[width=\columnwidth]{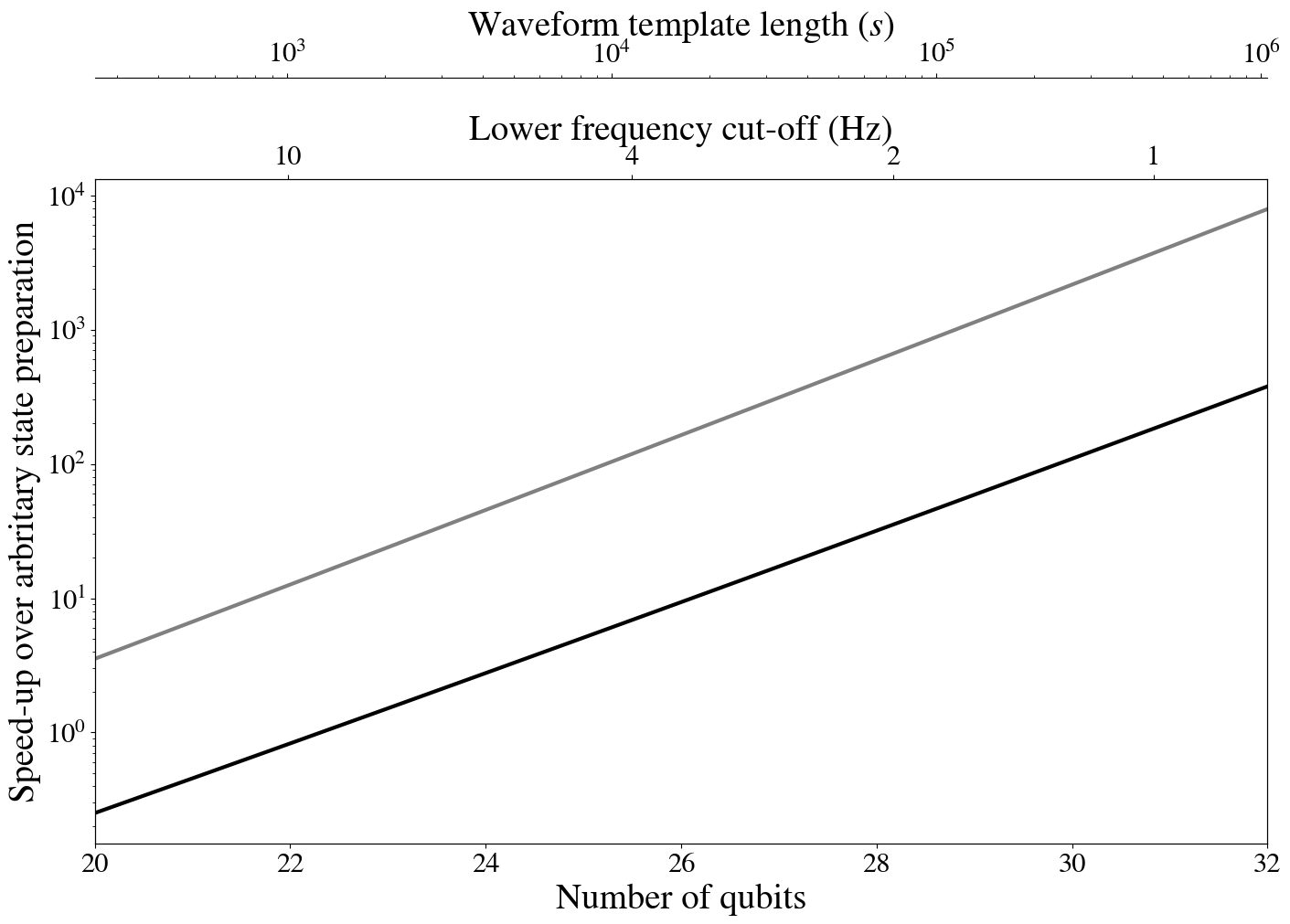}
\caption{Ratio of gate cost over duration when considering an arbitrary state preparation routine against the upper bound gate cost (see Appendix~\ref{sec:gatecost}) of preparing a insprial waveform using the method described in Sec.~\ref{sec:encode} with ${n_c=16}$, ${n_l=4}$ and ${n_a = n+n_c}$ using the Grover-Rudolph amplitude preparation routine plotted in black, and using a parameterised quantum circuit with ${L=100}$ in grey. Long duration waveforms that are required for third generation detectors, which probe the signals at low frequencies, can be produced by a factor of up to four orders of magnitude less gates.}
\label{fig:speedvlen}
\end{figure}
%\end{figure*}

We consider the case of waveforms required for the application of data analysis methods given future detector data, with higher sensitivities than the detectors that are currently in operation.
Detectors such as the Einstein Telescope will probe gravitational wave emissions at frequencies of as low as $1{-}7$\,Hz, where sources such as binary neutron star pairs can be apparent in the sensitive band for time periods of orders of days, a four orders of magnitude increase compared to current detectors~\cite{chan2018binary}.
The longer observation period allows for the orbital evolution of these systems to be studied in detail~\cite{maggiore2020science}. 
However, the increased sensitivity at lower frequencies for third generation detectors provide unique data analysis challenges and increases computational demand~\cite{bosi2011data}.
The longer duration signals require longer waveforms to perform coherent analyses, which is required to maximise sensitivity when considering a multi-detector network.
If a sampling rate of $4096$\,Hz is considered, the memory requirements to store a single binary neutron star inspiral waveform can extend into the gigabytes.
The majority of the matched filter signal-to-noise ratio accumulated by the Einstein telescope for binary neutron star inspirals will occur in the frequency region below $10$\,Hz~\cite{lenon2021eccentric}.
The matched filtering algorithm used to perform signal detection requires applying the fast Fourier transform to each of the $M$ templates which has a computational cost that scales as $O(MN\log N)$~\cite{allen2012findchirp}.
Therefore, given the size of the template bank $M$ must also grow to account for these low frequency signals, where generally the number of templates in the bank scales proportional to $f_{\text{min}}^{-8/3}$ to maintain a constant mismatch between potential signals and templates~\cite{owen1999matched}, longer duration signals will pose a computational challenge to the signal detection process~\cite{bosi2011data}.
This is further confounded by the need for a higher dimensionality template bank to account for spin and tidal deformability effects that increase the signal strength of higher-order terms of the waveform~\cite{forteza2018impact}.

By considering the method outlined in this work, the longest binary inspiral gravitational waveforms detectable to third generation detector, that require gigabytes of classical memory to store, can be stored on less than 32 qubits.
Fig.~\ref{fig:speedvlen} compares the gate cost of producing waveforms using this method when assuming ${n_c=16}$, ${n_l=4}$ and ${n_a = n+n_c}$ to the cost of using the arbitrary state preparation routine.
The black line shows the relative speed-up in terms of an arbitrary amplitude state preparation routine that requires $2^{n}$ controlled NOT gates~\cite{shende2005synthesis} when compared to the upper bound of controlled NOT gates of the analysis when using the Grover-Rudolph amplitude preparation routine specified in Eq.~\ref{equ:GRCX} of Appendix~\ref{sec:gatecost}.
While the overhead gate cost ensures that such a routine is inefficient for waveforms of duration less than $10^{3}$\,s, the method demonstrates a clear reduction in gate cost which can reach two orders of magnitude when waveforms of durations of up to ${\sim}10$ days are considered.
Waveforms of such durations are required for analysing low mass binary inspiral systems that are detectable by third generation detectors.

We note that the computational cost of preparing a waveform can be reduced from that described in Fig.~\ref{fig:speedvlen} at the expense of fidelity of the resulting amplitude encoded waveform by reducing $n_c$, $n_l$, $n_a$ or $m_b$.
Conversely the fidelity can be increased by increasing $n_c$ or $n_l$ with higher gate cost.
The solid grey line of Fig.~\ref{fig:speedvlen} depicts the relative speed-up when a trained parameterized quantum circuit like that of Eq.~\ref{equ:PQC} with ${L=100}$ layers of number of controlled NOT gates given in Eq.~\ref{equ:PQCCX} when compared to the $2^{n}$ controlled NOT gates of the arbitrary state preparation routine. 
The parameterised quantum circuit gains an order of magnitude reduction in controlled NOT gates when compared to the Grover-Rudolph algorithm amplitude preparation routine.

While the application of quantum algorithms to directly address the data analysis problems of third generation detectors is a task left for future work, we illustrate the power of the state preparation routine by showing the longest duration binary neutron star inspiral observable to third generation detectors can be amplitude encoded onto less than 32 qubits with the aid of 68 ancillary qubits.
This can be performed with an exponential speed-up over an arbitrary state preparation routine, a necessary condition to be met to qualify for any potential advantage of many quantum algorithms~\cite{aaronson2015read}.

\section{Conclusion}\label{sec:conc}

We provide an efficient routine for preparing amplitude encoded representations of the inspiral phase waveform of a gravitational wave signal emitted by the coalescence of a compact binary system.
The routine is based on the use of either the Grover-Rudolph algorithm or a generative hybrid classical-quantum machine learning method, and the ability to efficiently evaluate arithmetic functions through a piece-wise linear approximation.
We have demonstrated the application of this routine by simulating the preparation of a spinless $35~M_{\odot}$ and $30~M_{\odot}$ binary black-hole merger waveform in the frequency domain onto 6 qubits.
The resulting state obtains a fidelity of $0.995$ to the desired state when using the Grover-Rudolph algorithm method, which could further increase by increasing the number of available ancillary qubits, or assuming higher order polynomial approximations to evaluate the functions. 

A fidelity of $0.979$ was demonstrated when a generative adversarial network was used instead of the Grover-Rudolph algorithm, leading to a significant reduction in the gate count.
The resulting fidelity may be increased by training a larger parameterized circuit, however we demonstrate that training larger circuits requires more careful training regiments.
The consideration of other generative modelling schemes with more stable training may be beneficial, such as adopting a normalizing flow model~\cite{rezende2015variational}.
A further gate count reduction may be achieved by approximating the function evaluation step using a trained parameterised quantum circuit, rather than the current piece-wise function approximation that suffers from the high gate count required for arithmetic operations.

Amplitude encoding allows for exponentially larger waveforms to be encoded onto the given number of qubits in comparison to other encoding formats, which may allow for waveforms of sizes comparable to those used for classical analyses on qubit numbers obtainable on quantum processors in the near future.
We investigate this case when considering binary neutron star inspiral waveforms of lengths necessary for the analysis of signals from third generation detectors such as the Einstein Telescope.
Such detectors will encounter computational challenges when probing lower frequencies with higher sensitivity detectors.
We demonstrate that the state preparation method provides a greater speed-up over arbitrary state preparation methods the longer the waveform duration is, suggesting quantum advantages could be sought in this low frequency cut-off regime where the classical computational cost is greatest.

While we mainly pose the use of this routine for preparing states that represent gravitational waveforms, the same routine can be applied to prepare other functions of the form in Eq.~\ref{equ:hf}, including the merger and ringdown waveforms of gravitational wave binary merger events~\cite{ajith2008template}.
The adaption of this routine to perform state preparation of a combined inspiral, merger and ringdown waveform is left as future work.

Our work demonstrates that efficient encoding of gravitational waveforms into quantum states with a number of qubits that scales only logarithmically with the length of the waveform is possible. This extremely space-efficient encoding offers the tantalising possibility of near-term applications of quantum computational devices to gravitational wave astronomy. While the waveforms may be efficiently prepared, we note that this is not the case for encoding the detector data itself. This will require an arbritrary state preparation routine to encode and induce a computational cost that scales as $O(2^{n})$, which is unavoidable. Nonetheless, efficient encoding of template waveforms allows us to explore data analysis algorithms, with a potential speed-up in the training phase of machine learning approaches, or as a basis to construct measurements discriminating between signals in a measurement-based approach. We expect that our state preparation routine will find immediate applications in variational-based learning approaches to data analysis, as well as in the longer term to more sophisticated quantum protocols.

\section{Acknowledgements}

We are grateful for the support from the EPSRC under Grant No. EP/T001062/1.

\appendix

\section{Stationary phase approximation}\label{sec:SPA}

The waveforms that are represented analytically in the positive frequency-domain can be written:
\begin{equation}\label{equ:hf}
    \tilde{h}(f) \propto \tilde{A}(f) e^{i \Psi(f)},
\end{equation}
where $\tilde{h}(f)$ is the Fourier transform of real-valued $h(t)$: 
\begin{equation}\label{equ:TD}
    h(t) \propto \Re\left\{A(t)e^{-i\Phi(t)}\right\},
\end{equation}
such that $\tilde{h}(f) = \int h(t)e^{2\pi i ft}dt$, and $A$ and $\tilde{A}$ are real functions.
Commonly, $\tilde{h}(f)$ can be approximated from an analytical expression of Eq.~\ref{equ:TD} using the stationary phase approximation~\cite{thorne1987300} such that
\begin{equation}\label{equ:SPAA}
    \tilde{A}(f) \appropto A(t_0)\left(\frac{d^{2}\Phi(t_0)}{dt^{2}}\right)^{1/2}
\end{equation}
and
\begin{equation}\label{equ:SPAP}
    \Psi(f) \approx 2\pi f t_0 - \phi(f) - \frac{\pi}{4},
\end{equation}
where $t_0$ is the time at which $d\Phi(t)/dt|_{t_0}=2\pi f$, at which point $A(t)$ varies slowly.
The function $\phi(f)$ can be determined from $\Phi(t)$ given the frequency dependence on time through ${\Phi(t) = \int^{t}fdt'}$.

\section{Gravitational waves from compact binary inspirals}\label{sec:GWCBC}

Much of the following background is based on the work presented in Refs.~\cite{cutler1994gravitational,poisson1995gravitational}.
The gravitational waveform is entirely determined through general relativity. However, solutions to the non-linear Einstein field equations can only be found numerically, and analytical forms of the waveforms can only be approximated.
For compact binary systems, analytical waveforms are commonly determined using a post-Newtonian approximation in the near-field of the orbital system, while a post-Minkowskian approximation is made to the surrounding field~\cite{blanchet1995gravitational}.
Solutions are expanded about $(v/c)^{2}$, where ${v\ll c}$ is the  orbital velocity.
Taking only the leading-order terms leads to a `Newtonian' waveform which describes gravitational radiation produced solely from the change in the mass-quadrupole moment of the binary system, with frequency twice that of the orbital frequency~\cite{finn1993observing}. 
However, such a solution becomes inaccurate in systems with high mass ratios, component spins, or as the orbital velocity increases for systems with shorter separation where higher-order modes are excited.
This requires higher-order terms to be considered, leading to `post-Newtonian' waveforms.
Analytical waveforms are also often formed through fitting phenomenological models to sets of waveforms determined through numerical relativity~\cite{ajith2008template}.
For the remainder of this section, geometric units can be assumed such that ${G=c=1}$.

Gravitational radiation amplitudes depend on two independent polarization states denoted `$+$' and `$\times$', separated by an angle of $\pi/4$ from one another.
For a binary system of compact objects, these polarization amplitudes depend on the inclination $\iota$ of the system's angular momentum vector with respect to the line-of-sight, such that
\begin{align}
    h_+ =&  \frac{h_0}{2}(1+\cos^{2}\iota)\cos\Phi(t), \\
    h_\times =& h_0\cos\iota\sin\Phi(t).
\end{align}
The common amplitude $h_0$ is equal to
\begin{equation}
    h_0 \approx\frac{\mathcal{M}^{5/3}(2\pi f)^{2/3}}{D}.
\end{equation}
Here $\mathcal{M}$ is the chirp-mass, related to the total mass $M=m_1 + m_2$ and reduced mass $\eta=m_1m_2/(m_1 + m_2)$ of the system of component masses $m_{1,2}$ by $\mathcal{M} = \eta^{3/5} M^{2/5}$, and $D$ is the distance to the source.
To induce a time-dependent strain on a detector, the polarization amplitudes couple to the detector's response to each polarization $F_{+,\times}$ and combine linearly:
\begin{equation}
    h(t) = F_+h_+ + F_\times h_\times = Q h_0\cos(\Phi(t) - \Phi_0).
\end{equation}
This can be rewritten as
\begin{equation}\label{equ:TDGW}
    h(t) = Q h_0\cos(\Phi(t) - \Phi_0),
\end{equation}
where ${Q = (A_+^2 + A_\times^2)^{1/2}}$, and ${\Phi_0 = \arctan(A_\times/A_+)}$, given ${A_+ = F_+(1+\cos^2\iota)/2}$ and ${A_\times = F_\times \cos\iota}$.
The expression in Eq.~\ref{equ:TDGW} is in the form of Eq.~\ref{equ:TD}.

The spin of the binary system is parameterised by the spin-orbit parameter:
\begin{equation}
    \beta = \frac{1}{12}\sum^{2}_{i=1}\boldsymbol{L}\cdot\boldsymbol{\chi}_i[113(m_i/M)^{2}+75\eta],
\end{equation}
and spin-spin parameter

\begin{equation}
    \sigma = \frac{\eta}{48}\left(721(\boldsymbol{L}\cdot\boldsymbol{\chi}_1\boldsymbol{L}\cdot\boldsymbol{\chi}_2) -247(\boldsymbol{\chi}_1\cdot\boldsymbol{\chi}_{2})\right).
\end{equation}
Here, $\boldsymbol{\chi}_{1,2}=\boldsymbol{S}_{1,2}/m_{1,2}^{2}$ where $\boldsymbol{S}_{1,2}$ is the spin angular momentum of bodies $1$ and $2$, which is compared against the orbital angular momentum of the system $\boldsymbol{L}$.

The time-frequency dependence to second post-Newtonian order is
\begin{widetext}\label{equ:dfdt}
\begin{multline}
    \frac{df}{dt} = \frac{96}{5\pi\mathcal{M}^{2}}(\pi\mathcal{M}f)^{11/3}\left[1-\left(\frac{743}{336}+\frac{11}{4}\eta\right)(\pi Mf)^{2/3} \right. \\ \left. + (4\pi - \beta)(\pi M f) + \left(\frac{34103}{18144} + \frac{13661}{2016}\eta + \frac{59}{18}\eta^{2} + \sigma\right)(\pi Mf)^{4/3}\right].
\end{multline}
\end{widetext}

From Eq.~\ref{equ:dfdt}, the frequency dependence on time can be calculated, giving the phase as $\Phi(t) = \int^{t}fdt'$ from which Eq.~\ref{equ:Psif2N} is calculated from Eq.~\ref{equ:SPAP}.
The time-frequency relation can be inverted allowing for $\phi(f) = \Phi(t(f))$ to be determined, resulting in the expression for the frequency dependent amplitude of Eq.~\ref{equ:Af} given Eq.~\ref{equ:SPAA}.

\section{Fixed-point binary encoding}

Our routine to prepare the state of Eq.~\ref{equ:targ} requires binary strings to be encoded in the computational basis using \textit{signed-magnitude} (and unsigned-magnitude) and \textit{two's-complement} representations, which we detail here.

We represent signed binary numbers in a bit-strings of $n$ bits with the ordering:
\[x = \underbrace{x_{n-1}}_{\text{Sign bit}} \underbrace{x_{n-2}\dots x_{n-n_{\text{int}}-1}}_{n_{\text{int}}\text{ integer bits}} \underbrace{x_{p-1}\dots x_0}_{p\text{ precision bits}},\]
so that the sign of the number is represented in the leading order bit, the integer part in the following $n_{\text{int}}$ bits, and the fraction part in the final $p$ precision bits.
Note that while $n_{\text{int}}$ and $p$ must follow the condition that $n = n_{\text{int}}+p+1$, the values are not restricted to positive numbers, allowing for the precision $2^{-p}\ge 1$ or the upper bound to $x$ that the bit-string can represent to be $2^{n_{\text{int}}}\le 1$.

\subsection{Unsigned-magnitude representation}\label{sec:mrep}

When only the magnitude of $x$ is required, the sign bit is dropped and the representation of the number is simply:

\[x = \sum^{n-1}_{i=0}x_{i}2^{i-p}.\]

\subsection{Signed-magnitude representation}\label{sec:smrep}

Positive and negative numbers of equal magnitude are related by flipping the leading order bit (the sign bit) so that:

\[x = (-1)^{x_{n-1}}\sum^{n-2}_{i=0}x_{i}2^{i-p}.\]

Note that this representation includes both a positive and negative value for zero.

\subsection{Two's-complement representation}\label{sec:2comprep}

Negative numbers are related to positive values of equal magnitude by a bit-flip of all bits and incrementing the result by the least significant bit (equivalent to the addition of $2^{-p}$):

\[x = -x_{n-1}2^{n_{\text{int}}} + \sum^{n-2}_{i=0}x_{i}2^{i-p}.\]

\section{Fixed-point multiplication through the quantum Fourier transform}\label{sec:mult}

%Multiplication of fixed-point numbers can be performed by applying the qua to the output register before, and the inverse after, for both the multiplication and the addition operations.
The multiplication of fixed-point numbers can be carried out without ancillary qubits through the use of the quantum Fourier transform.
The multiplicand is stored in state $\ket{a}$ and the multiplier in $\ket{b}$.
A third, initially empty, output register $\ket{0}_o$ is required to store the product such that
\begin{align*} 
\ket{a}\ket{b}\ket{0}_o &\xrightarrow{\text{Mult}} \ket{a}\ket{b}\ket{ab}_o.
\end{align*}
Like classical multiplication algorithms, the multiplication is performed by a series of additions of $\ket{a}$ onto $\ket{0}_o$ of a number depending on $\ket{b}$.
However, here we perform the additions to $\ket{0}_o$ by first performing the quantum Fourier transform across the register.
The addition of $\ket{a}$ onto the output register while in the conjugate basis space is then a series of controlled rotation operations.
The details of how these additions are performed is given in Ref.~\cite{ruiz2017quantum}, which is expanded to multiplication by further controlling the rotational addition operations on the multiplier $\ket{b}$. 
Similar to \cite{haner2018optimizing}, we simplify the multiplication by asserting that the multiplicand is positive, and the result is processed conditioned on the most significant qubit; if the multiplier is negative (the most significant qubit is $\ket{1}$), the output is bit-flipped and incremented by 1 to correspond with negative values in the two's complement notation.
As the multiplication product is encoded in the conjugate basis in two's-complement representation, the following addition operation of Fig.~\ref{fig:Pgate} is applied independently of the product's sign.

\section{Gate cost}\label{sec:gatecost}

An upper bound to the number of controlled $X$ gates can be assigned to each of the operations used in the the state preparation routine. 
This upper bound could be reduced by optimising the implementation of each of the operations or by introducing additional ancillary qubits.
However, we provide the upper bound when assuming the form of each operation as used to perform the quantum simulation described in Sec.~\ref{sec:GWI}.
For the multiplication of a register of length $n_1$ on one of $n_2$ where the result is stored on a register of length $n_1 + n_2$, the upper bound to the number of controlled $X$ gates is:
\begin{align}
    C_{\text{Mult}}(n_1,n_2) =& 8(n_1 + n_2)n_{2}(n_1 - 1) \nonumber\\
    &+ 20(n_1+n_2)^{2} \nonumber\\
    &  - 13(n_1 + n_2).
\end{align}
Similarly the addition of registers of length $n_1$ and $n_2$, where the result is stored on the second register requires:
\begin{equation}
    C_{\text{Add}}(n_1, n_2) = 2n_1n_2 + 2n_2(n_2-1).
\end{equation}

The label operation when utilizing the quantum Fourier transform is bound by:
\begin{equation}
    C_{\text{Label}}(n, n_l) = 2n_l(n_l-1) + 2^{n_l+1}(6n + n_l + 1).
\end{equation}
The inputting of the piece-wise function coefficients requires:
\begin{equation}
    C_{\text{X}}(n_c, n_l) = n_c2^{n_l}C_{C^{\otimes n}X}(n_l),
\end{equation}
where $C_{C^{\otimes n}X}$ is the number of controlled $X$ gates required to perform a $n$-controlled $X$ operation.
Without the use of ancillary qubits $C_{C^{\otimes n}X}(n_l)=2n_l^2 - 6n_l + 5$~\cite{saeedi2013linear}, otherwise $C_{C^{\otimes n}X}(n_l)=20(n_l-2)$ with the use of $n-2$ ancillary qubits~\cite{barenco1995elementary} or $12(n_l - 1) + 1$ with $n-1$ ancillary qubits~\cite{nielsen2002quantum}.

The linear piece-wise function (as shown in Fig.~\ref{fig:Pgate}) is then constrained by:
\begin{align}
    C_{\text{LPF}}(n, n_c, n_l) =& C_{\text{Mult}}(n_c,n)+ C_{\text{Add}}(n_c, n_c+n)\nonumber\\
    &+ C_{\text{Label}}(n, n_l) + 3C_{\text{X}}(n_c, n_l)
\end{align}

Then the upper bound on the combined amplitude and phase preparation steps is placed when assuming $m_a=1$ and $m_b=n$:
\begin{widetext}
\begin{align}\label{equ:GRCX}
    C_{\text{GW}}(n, n_c, n_l) =& 2\sum_{m=1}^{n}C_{\text{LPF}}(m, n_c, n_l) + 2(n+n_c(n-1)).
\end{align}
\end{widetext}

If a trained parameterized quantum circuit is used, the amplitude preparation step can be reduced to $L(n-1)$ controlled $X$ gates, reducing the total count to:
\begin{align}\label{equ:PQCCX}
    C_{\text{GW}}(n, n_c, n_l) =& 2C_{\text{LPF}}(n, n_c, n_l) + L(n-1).
\end{align}

\bibliographystyle{unsrt}
\bibliography{combined} 

\end{document}